\documentclass[aps,prb,twocolumn,superscriptaddress]{revtex4-1}

\usepackage{graphicx}
\usepackage{amsmath}
\usepackage{amssymb,mathtools}
\usepackage{braket}
\usepackage{color}
\usepackage[normalem]{ulem}
\usepackage[dvipsnames]{xcolor}
\usepackage{siunitx}
\usepackage{mathrsfs}
\usepackage{xpatch} % For bibliographic count 
\DeclareGraphicsExtensions{.eps}

\newcommand*{\rom}[1]{\expandafter\@slowromancap\romannumeral #1@}

\begin{document}

\title{Steering of Quantum Walks through Coherent Control of High-dimensional Bi-photon Quantum Frequency Combs with Tunable State Entropies}
    
\author{Raktim Haldar}
	\email{raktim.haldar@iop.uni-hannover.de}
	\affiliation{Institute of Photonics, Leibniz University Hannover, Nienburger Stra\ss e 17, D-30167 Hannover, Germany}
    \affiliation{Hannover Centre for Optical Technologies, Leibniz University Hannover, Nienburger Stra\ss e 17, D-30167 Hannover, Germany}
   \affiliation{Cluster of Excellence PhoenixD (Photonic, Optics, and Engineering – Innovation Across Disciplines), Leibniz University Hannover, Hannover, Germany}
	
\author{Robert Johanning}
	%\email{}
	\affiliation{Institute of Photonics, Leibniz University Hannover, Nienburger Stra\ss e 17, D-30167 Hannover, Germany}
    \affiliation{Hannover Centre for Optical Technologies, Leibniz University Hannover, Nienburger Stra\ss e 17, D-30167 Hannover, Germany}
    \affiliation{Cluster of Excellence PhoenixD (Photonic, Optics, and Engineering – Innovation Across Disciplines), Leibniz University Hannover, Hannover, Germany}

   \author{Philip R\"ubeling}
	%\email{}
	\affiliation{Institute of Photonics, Leibniz University Hannover, Nienburger Stra\ss e 17, D-30167 Hannover, Germany}
    \affiliation{Hannover Centre for Optical Technologies, Leibniz University Hannover, Nienburger Stra\ss e 17, D-30167 Hannover, Germany}
    \affiliation{Cluster of Excellence PhoenixD (Photonic, Optics, and Engineering – Innovation Across Disciplines), Leibniz University Hannover, Hannover, Germany}

\author{Anahita Khodadad Kashi}
	%\email{}
	\affiliation{Institute of Photonics, Leibniz University Hannover, Nienburger Stra\ss e 17, D-30167 Hannover, Germany}
    \affiliation{Hannover Centre for Optical Technologies, Leibniz University Hannover, Nienburger Stra\ss e 17, D-30167 Hannover, Germany}
   \affiliation{Cluster of Excellence PhoenixD (Photonic, Optics, and Engineering – Innovation Across Disciplines), Leibniz University Hannover, Hannover, Germany}

\author{Thomas B{\ae}kkegaard}
	\affiliation{Institute of Photonics, Leibniz University Hannover, Nienburger Stra\ss e 17, D-30167 Hannover, Germany}
    \affiliation{Hannover Centre for Optical Technologies, Leibniz University Hannover, Nienburger Stra\ss e 17, D-30167 Hannover, Germany}
    \affiliation{Kvantify APS, DK-2300, Copenhagen S, Denmark}

\author{Surajit Bose}
	%\email{}
	\affiliation{Institute of Photonics, Leibniz University Hannover, Nienburger Stra\ss e 17, D-30167 Hannover, Germany}
    \affiliation{Hannover Centre for Optical Technologies, Leibniz University Hannover, Nienburger Stra\ss e 17, D-30167 Hannover, Germany}
   \affiliation{Cluster of Excellence PhoenixD (Photonic, Optics, and Engineering – Innovation Across Disciplines), Leibniz University Hannover, Hannover, Germany}
	
\author{Nikolaj Thomas Zinner}
     \affiliation{Kvantify APS, DK-2300, Copenhagen S, Denmark}
     \affiliation{Department of Physics and Astronomoy, Aarhus University, Aarhus C DK-8000, Denmark}

\author{Michael Kues}
	\email{michael.kues@iop.uni-hannover.de}
	\affiliation{Institute of Photonics, Leibniz University Hannover, Nienburger Stra\ss e 17, D-30167 Hannover, Germany}
    \affiliation{Hannover Centre for Optical Technologies, Leibniz University Hannover, Nienburger Stra\ss e 17, D-30167 Hannover, Germany}
   \affiliation{Cluster of Excellence PhoenixD (Photonic, Optics, and Engineering – Innovation Across Disciplines), Leibniz University Hannover, Hannover, Germany}	

\begin{abstract}{Quantum walks are central to a wide range of applications such as quantum search, quantum information processing, and entanglement transport. Gaining control over the duration and the direction of quantum walks (QWs) is crucial to implementing dedicated processing. However, in current systems, it is cumbersome to achieve in a scalable format. High-dimensional quantum states, encoded in the photons' frequency degree of freedom in on-chip devices are great assets for the scalable generation and reliable manipulation of large-scale complex quantum systems. These states, viz. quantum frequency combs (QFCs) accommodating huge information in a single spatial mode, are intrinsically noise tolerant, and suitable for transmission through optical fibers, thereby promising to revolutionize quantum technologies. Existing literature aimed to generate maximally entangled QFCs excited from continuous-wave lasers either from nonlinear microcavities or from waveguides with the help of filter arrays. QWs with flexible depth/duration have been lately demonstrated from such QFCs. In this work, instead of maximally-entangled QFCs, we generate high-dimensional quantum photonic states with tunable entropies from periodically poled lithium niobate waveguides by exploiting a novel pulsed excitation and filtering scheme. We confirm the generation of QFCs with normalized entropies from $\sim 0.35$ to $1$ by performing quantum state tomography with high fidelities. These states can be an excellent testbed for several quantum computation and communication protocols in nonideal scenarios and enable artificial neural networks to classify unknown quantum states. Further, we experimentally demonstrate the steering and coherent control of the directionality of QWs initiated from such QFCs with tunable entropies. Our findings offer a new control mechanism for QWs as well as novel modification means for joint probability distributions.\\

}
\end{abstract}

\maketitle

\section{Introduction}
 Quantum information processing (QIP) aided by photonics is a promising approach due to the availability of proficient manipulation tools at room temperature and the robustness to decoherence \cite{zhong2020quantum, o2009photonic}. High-dimensional quantum systems, i.e., multi-level systems employing qudits as basic computational units are inherently more resilient to noise, fault-tolerant, and by leveraging the elegance of high-dimensional quantum algorithms enhance the performance of quantum computers \cite{erhard2020advances, sheridan2010security, cerf2002security}. High-dimensional quantum states exploiting the photons’ frequency degrees of freedom, so-called quantum frequency combs (QFCs), can be generated on-chip, manipulated through off-the-shelf telecommunication components, carried over long distances via commercial optical fibers, and can further help in realizing large-scale complex quantum states.  Thus, QFCs constitute highly coveted scalable resources for QIP and quantum communication (QC) \cite{kues2019quantum, kues2017chip, imany201850, francesconi2020engineering, reimer2019high}. QFCs that have been previously demonstrated in on-chip optical microresonators (MRs) through spontaneous four-wave mixing (SFWM) \cite{kues2017chip, imany201850} or by judiciously carving the continuous-wave spontaneous parametric down-conversion (SPDC) spectra from periodically poled lithium niobate (PPLN) waveguides, form maximally entangled $d$-dimensional Bell-states \cite{imany2018characterization}. 
%----------------------------------Steering of QW paper Fig. 1.----------------------------------
\begin{figure*}[t]
    \centering
    \includegraphics[width=0.8\textwidth,trim=0mm 0mm 0mm 0mm, clip]{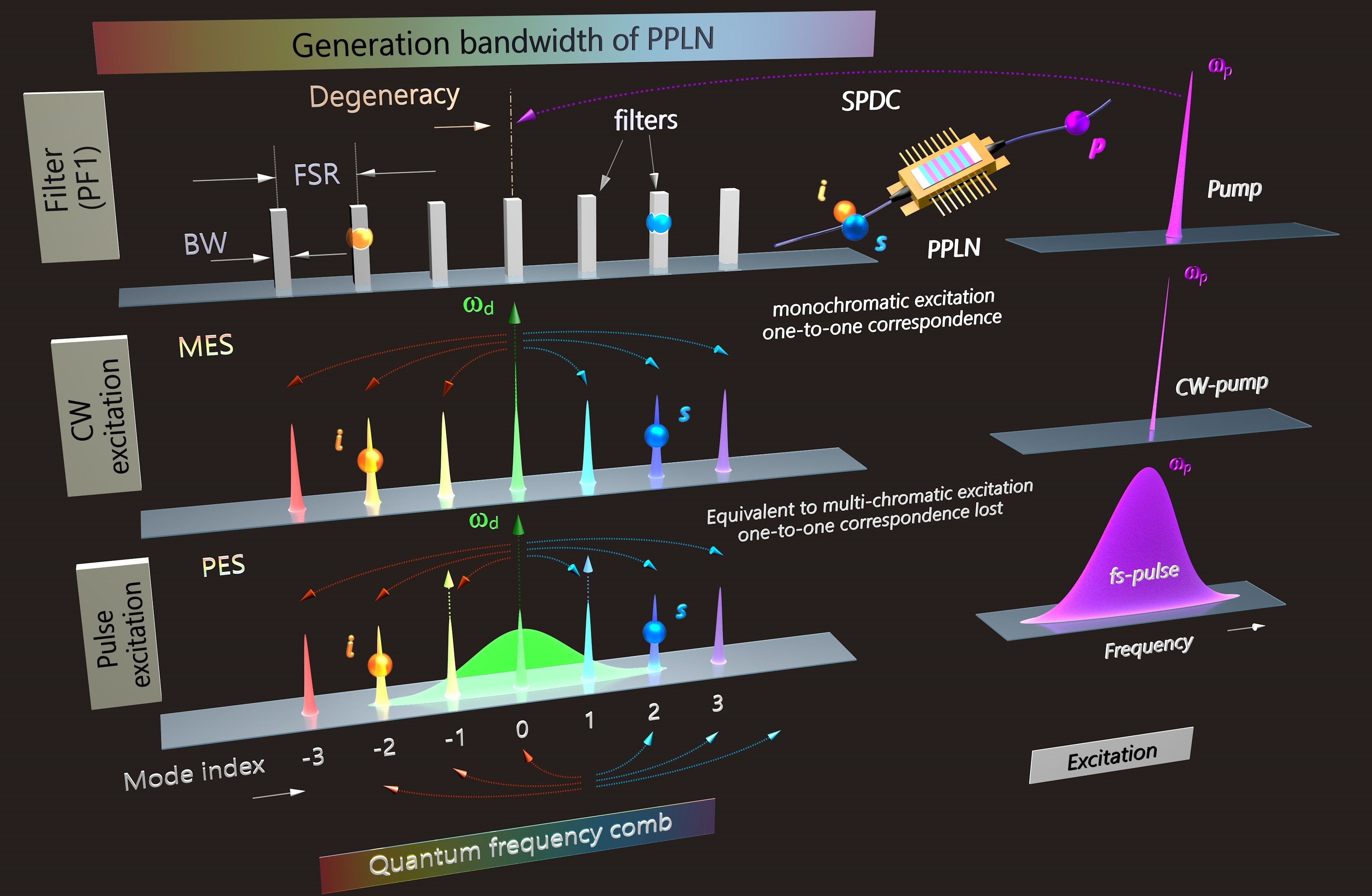}
    \caption{
    \textbf{Generation scheme of non-maximally entangled quantum frequency comb (QFC) from a femtosecond pulse excitation.} A QFC can be generated through spontaneous parametric down-conversion (SPDC) by pumping a periodically poled lithium niobite (PPLN) waveguide and by using an equidistant filter array. When the waveguide is pumped with a CW-excitation of angular frequency ($\omega_{\mathrm{p}}$), the generated signal ($s$) and idler ($i$) around the degenerate angular frequency ($\omega_{\mathrm{d}}$) have a one-to-one correspondence, which means that the detected frequency of the signal photon explicitly dictates the frequency of the corresponding idler photon. Therefore, the QFC is maximally entangled. On the other hand, if the waveguide is excited through a broadband pulse (center frequency $\omega_{\mathrm{p}}$), the one-to-one correspondence is lost and the QFC becomes non-maximally entangled. This pulse excitation scheme is similar to the multi-chromatic excitation of an optical resonator, which also yields non-maximally entangled QFCs.
    }
    \label{fig: QFC scheme}
\end{figure*}
%----------------------------------Steering of QW paper Fig. 1.----------------------------------
Ideally, maximally entangled quantum states (MES) are regarded as the most crucial resources for QIP and QCs. However, in practice, separable, weakly entangled, and non-maximally entangled quantum states (NMES) or the partially entangled states (PES) are extremely useful tools for quantum computing in dissipative environments \cite{dakic2012quantum}. In the presence of decoherence, it is challenging to retain the quality of a MES intact, whereas, state preparation, manipulation, and maintenance of the entanglement through a noisy channel are easier for non-maximally entangled pure and mixed states \cite{dakic2012quantum}. For instance, without filtering and entanglement concentration, conventional teleportation protocols with MES shared as a resource between the sender and the receiver cannot guarantee unit fidelity for an unknown state \cite{agrawal2002probabilistic, bandyopadhyay2012optimal}. In the presence of amplitude damping throughout an imperfect channel, utilizing partially entangled resources are advantageous for quantum correlation distribution and storage over MES \cite{wang2015nonmaximally, chen2021perfect}. Similarly, the entanglement lower bound ensuring the optimal fidelity for local cloning of a pure bi-partite system, and the roles of quantum correlation on the fidelity of cloning and deletion are theoretically established \cite{demkowicz2006usefulness, sazim2015complementarity}. Non-maximally entangled states are known to reduce the required detector efficiencies for loophole-free tests of Bell inequalities \cite{white1999nonmaximally, giustina2015significant}. Besides increasing the accessibility of Hilbert space, such non-maximally entangled states can be implemented in nonlocality (e.g., Hardy's proof) tests without inequalities, thereby enabling us to answer fundamental questions of quantum mechanics \cite{white1999nonmaximally}. Recently, PESs are employed to examine the effectiveness of entanglement concentration, distillation protocols, and random number generation schemes in practical scenarios \cite{zhao2016entanglement, vaziri2003concentration, gomez2019experimental}. Therefore, we expect that high-dimensional quantum photonic states with tunable entropies suitable for long-haul communications would also offer excellent testbeds for other QIP protocols such as teleportation, cloning, deletion, purification, error-correction, and would be suitable for noisy-intermediate scale quantum (NISQ) technologies \cite{preskill2018quantum}. Apart from that, for quantum algorithms, such as the high-dimensional Deutsch algorithm, amplitude encoding for quantum algorithms \cite{wang2020qudits, nakaji2021approximate}, or for producing pure heralded photons, often high-dimensional product (separable) states are required. These are not trivial to realize, as it requires broad energy transitions with a bandwidth that is accessible with optical filters \cite{mosley2008heralded, soller2011high}.  
 
Moreover, quantifying the entropy of an entangled resource and determining its efficacy for a certain QIP or a cryptographic protocol could be of paramount interest. Nevertheless, unlike energy, entanglement does not correspond to any observable and there is no straightforward experimental procedure to obtain the entanglement of an unknown quantum system. Especially measuring entanglement of a strongly correlated many-body quantum system remains an intractable problem to date \cite{islam2015measuring}. Therefore, it is necessary to generate quantum states with a range of entropies, for instance as resources to train neural networks, able to directly classify unknown quantum states and their suitability in QC channels \cite{harney2020entanglement}. 
 
Recently, we proposed pumping schemes in an MR \cite{haldar2021multi} based QFC that facilitate the excitation of multiple anti-diagonal coincidence lines that reduces the frequency correlation, thereby allowing us to tune the entropy of the QFC. In this work, we demonstrate a novel pumping and filtering scheme \cite{haldar2021high} in PPLN waveguide-based QFC to maneuver its entropy, by tailoring its joint spectral intensity (JSI). Further, gaining control over the direction of quantum walk (QW), although very critical to QIP and quantum information transport \cite{albertini2012controllability}, has been indeed difficult \cite{katayama2020floquet, novo2021floquet}, and often impossible due to its inherent stochastic nature. Unidirectional one and two-dimensional QW from separable two-particle states have been shown theoretically \cite{omar2006quantum}. Nevertheless, the experimental demonstration of obtaining a definitive control over the directions of QWs in photonic systems is still scarce. To address this issue, we study quantum walks initiated from such high-dimensional quantum states with variable entropies \cite{haldar2021steering} and demonstrate control. 

Quantum superpositions enable QWs \cite{aharonov1993quantum, preiss2015strongly} to potentially speedup certain computational tasks such as database searches, tests of graph isomorphism, ranking nodes in a network \cite{schreiber2010photons}, quantum many-body simulations \cite{preiss2015strongly}, boson sampling \cite{qiang2016efficient}, universal quantum computing \cite{childs2009universal, lovett2010universal}, and quantum state preparation \cite{zhang2016creating}. Entanglement generation, localization, and quantum information transport through QW even find applications in exotic fields of studies, notably, in explaining the energy transfer mechanism within photosynthesis \cite{mohseni2008environment}, in neural network \cite{schuld2014quantum}, for topology identification \cite{ming2019quantum}, and in neuroscience \cite{hameroff2014quantum}. Being robust and immune to decoherence at room temperature, QW realized in photonic platforms are advantageous over other platforms such as cold atom, Bose-Einstein condensates (BEC), optical lattices, trapped ions, etc.  \cite{peruzzo2010quantum}. However, QW implementing spatial \cite{peruzzo2010quantum}, polarization \cite{schreiber2010photons}, angular momentum \cite{cardano2017detection} degrees of freedom of photon either require large overhead to alter the depth of the QW \cite{schreiber2010photons} or necessitate modifying the physical layout to attain the tunability of the duration of QW \cite{peruzzo2010quantum, sansoni2012two}. Recently, QWs exhibiting enhanced ballistic transport (bosonic) or strong energy confinement (fermionic) have been demonstrated \cite{imany2020probing} using high-dimensional bi-photon quantum frequency combs (QFCs) \cite{imany2020probing}, which do not require any change of the device arrangement. However, no control over the directions of the demonstrated QWs \cite{imany2020probing} could be achieved, which were initiated from the maximally entangled states. Recently, Floquet engineered discrete- and continuous-time QWs and their control have been reported numerically by using time-dependent coins \cite{katayama2020floquet}, and by tweaking the node-coupling coefficients \cite{novo2021floquet}. The role of space-dependent coins \cite{panahiyan2018controlling}, and initial conditions \cite{de2010tailoring} on QWs are also studied extensively. Lately, the directionality of QW in BEC has been observed \cite{dadras2018quantum, weiss2015steering}. Quantum photonic states having tunable entropies extend our accessibility to the Hilbert space. Consequently, richer dynamics of QWs instigated from such states are expected yet have not been observed to date. For the first time to the best of our knowledge, here we experimentally demonstrate the coherent control of the direction and steering of quantum walk initiated from a high-dimensional bi-photon quantum frequency comb with tunable state entropies leading to a completely new paradigm of QWs. Procuring precise control over a nonclassical stochastic process involving a high-dimensional Hilbert space may have immense implications in, e.g., quantum search, transport of quantum information, and atomic interference.    

\section{Description and characteristics of non-maximally entangled QFCs}
Biphoton quantum frequency combs providing a route to generate complex high-dimensional states \cite{reimer2019high} can play a major role in discrete-variable photonic quantum computing. They are either generated by monochromatically exciting one of the resonating modes of a microresonator (MR) or by pumping a nonlinear waveguide typically below the parametric threshold \cite{kues2019quantum} with the help of an array of equidistant optical filters. We adopted the latter approach at the moderate expense of reduced brightness and second-order correlation ($g^{(2)}$) because of the increased design flexibility of choosing the QFC free-spectral range (FSR), also necessary to synchronize with the RF-driving frequency to perform the quantum state tomography (QST). 

A scheme for the generation of non-maximally entangled QFCs from pulse excitation is shown in Fig. \ref{fig: QFC scheme}. A  programmable filter (PF1) configuration is used to discretize the SPDC spectra from a femtosecond (fs) laser-driven PPLN-waveguide to create a bi-photon QFC \cite{imany201850}. The distance between the adjacent bandpass filters (BPFs) defines the free spectral range (FSR) of the QFC. The femtosecond (fs)-laser allows producing a broad single-mode frequency bandwidth for each photon with respect to CW-pumping \cite{imany2020probing, khodadad2021spectral, kashi2022frequency}. Together with a special filter, this may result in multiple anti-diagonal lines in the JSI. Due to the presence of multiple lines, one particular idler mode is spectrally connected to several signal modes (instead of one, as it would be in a maximally entangled state). This effectively reduces the entanglement of the system as explained in Fig. \ref{fig: QFC scheme}. One can create a variety of JSIs with several frequency anticorrelation line configurations by changing the spectral profile of the excitation or by altering the FSRs. To simulate such versatile JSIs, we develop a mathematical model (see Supplementary Information), where we assume that the BPF corresponding to the $0$-th (central) mode is placed matching with the center of the quasi-phase-matching (QPM) bandwidth (i.e., the degenerate angular frequency $(f_{\mathrm{d}})$ $=$ pump-frequency $(f_\mathrm{p})/2$) of the PPLN.

Initially we model the corresponding JSI of the QFC assuming there exists an equal number of frequency anti-correlation lines (i.e., symmetrical) surrounding the central antidiagonal. The quantum-state representing the $p-$th adjacent anti-diagonals with respect to the central anti-diagonal ($p = 0$) of such QFC having $N$ number of total frequency bins can be approximated by,
\begin{equation}
\footnotesize
{
\left| {{\psi }_{q}} \right\rangle =\frac{1}{\sqrt{\frac{\left( N-1 \right)}{2}-\left( \left| q \right|-1 \right)}}\sum\limits_{k=\left| q \right|}^{{(N-1)}/{2}\;}{\left\{ \begin{array}{*{35}{l}}
   {{\left| k+q \right\rangle }_{s}}{{\left| -k \right\rangle }_{i}}; & q<0  \\
   {{\left| k \right\rangle }_{s}}{{\left| -k \right\rangle }_{i}}; & q=0  \\
   {{\left| k \right\rangle }_{s}}{{\left| -\left( k-q \right) \right\rangle }_{i}}; & q>0  \\
\end{array} \right.}
}
\label{eq: Psi_expression}
\end{equation}

\begin{figure*}[htbp]
    \centering
    \includegraphics[width=0.85\textwidth,trim=0mm 0mm 0mm 0mm]{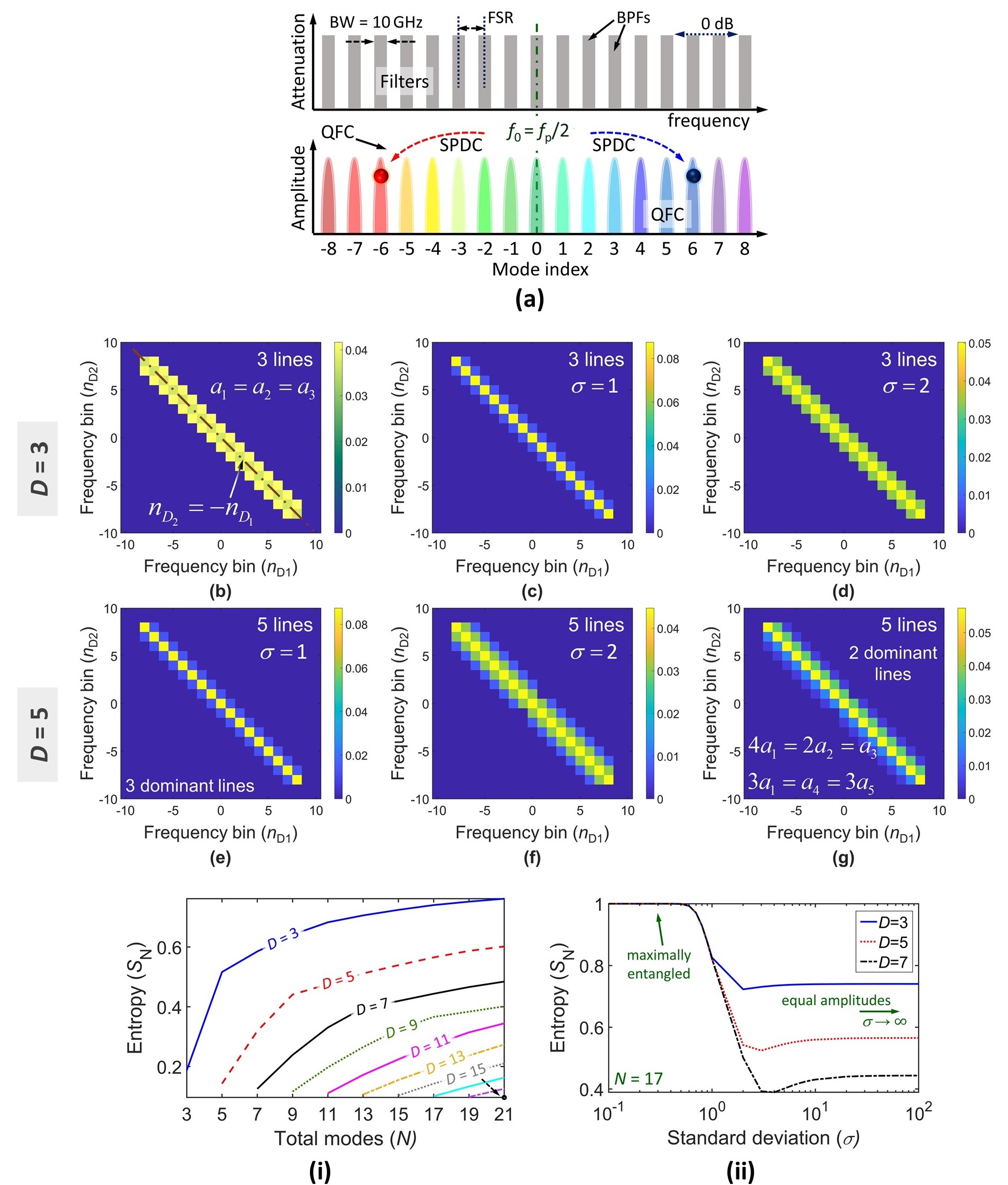}
    \caption{
    \textbf{Joint spectral intensities of non-maximally entangled QFCs and the dependency of their entropies on different parameters (a)} Filter configuration to generate a bi-photon QFC with $N = 17$ modes. Simulated joint spectral intensity (JSI) with three frequency anti-correlation lines \textbf{(b)} having equal amplitudes, and having Gaussian distribution of standard deviations \textbf{(c)} $\sigma = 1$, and \textbf{(d)} $\sigma = 2$. Five frequency anti-correlation lines having Gaussian distribution of standard deviations \textbf{(e)} $\sigma = 1$, \textbf{(f)} $\sigma = 2$, and \textbf{(g)} asymmetric distribution with two dominant lines. \textbf{(i)} Normalized Von Neumann entanglement entropy ($S_\mathrm{N}$) as a function of the total number of QFC-modes $N$ with varying JSI-diagonals $D$s in the JSI when all the diagonals are equiprobable (i.e., $a_\mathrm{q} = 1$).  \textbf{(ii)} $S_\mathrm{N}$ with the standard deviation ($\sigma$) when the JSI follows the normal distribution. BW: bandwidth.}
    \label{fig: JSI and entropy}
\end{figure*}

Where, $q < 0$, $q = 0$, $q > 0$ represent the lower, central, and upper diagonals, respectively. The complete state-vector of the QFC $\left| {{\psi }_{\mathrm{QFC}}} \right\rangle$ can be written as the summation of each normalized diagonal element weighted by the complex diagonal contribution $a_\mathrm{q}$. Fig. \ref{fig: JSI and entropy} (b)-(d) show an examples of the JSIs calculated from Eq. (\ref{eq: Psi_expression}), where the amplitudes of the three diagonals are (b) equiprobable (i.e., $a_1 = a_2 = a_3$), and following a Gaussian distribution with standard deviations (c) $\sigma = 1$, and (d) $\sigma = 2$, respectively. Thereafter, we simulate QFCs with five frequency anticorrelation lines in the JSI with Gaussian distribution having standard deviations (e) $\sigma = 1$, (f) $\sigma = 2$, and (g) asymmetric distribution with respect to the central antidiagonal, where two out of five lines are dominant. By diagonalizing the reduced density matrix of rank $r$ obtained from Eq. (\ref{eq: Psi_expression}), we can compute the von Neumann entropy ${{S}_{\mathrm{A}}}=-\sum{{{\lambda }_{\mathrm{m}}}}{{\log }_{2}}\left( {{\lambda }_{\mathrm{m}}} \right)$. Since entropy increases with dimension, we introduced the normalized entropy ${{S}_{\text{N}}}=-\sum{{{\lambda }_{\text{m}}}}{{\log }_{\mathrm{r}}}\left( {{\lambda }_{\text{m}}} \right)$, where $\lambda_m$ are the eigenvalues of $\hat{\rho }$. In Fig. \ref{fig: JSI and entropy} (i), we plot $S_\mathrm{N}$ with respect to the total number of modes $N$ while varying the number of diagonals $D$ in the JSI from $3$ to $N$. Note that, if the frequency modes are equiprobable, for $D = 1$, the QFC is maximally entangled with $S_\mathrm{N} = 1$. From Fig. \ref{fig: JSI and entropy} (i), it can also be seen that with an increasing number of diagonals, the entropy of the QFC reduces. The entropy of the QFC increases with increasing $N$. We further study the effect of the JSI spectral bandwidth, i.e., the combined effect of the pump spectral distribution and the QPM bandwidth on entropy. We assume that the amplitudes $a_\mathrm{q}$ of the JSI-diagonals follow a Gaussian distribution with standard deviation $\sigma$. As depicted in Fig. \ref{fig: JSI and entropy} (ii), for small values of $\sigma$, there is effectively a single JSI-line corresponding to the maximally entangled QFC, whereas, $\sigma \to \infty$ creates the QFC with equal amplitudes, i.e., for $\forall a_\mathrm{q} = 1$. Note that, we can also omit the degenerate mode (central mode) of the QFC by the programmable filter. In that case, the rank of the reduced density matrix $r$ coincides with the dimension ($d$) of the QFC (mathematical details can be found in the Supplementary Materials).

%----------------------------------Steering paper Fig. 2.----------------------------------
%----------------------------------Steering of QW paper Fig. 2.----------------------------------

\section{Experimental demonstration of QFC with tunable entropies}
\begin{figure*}[t]
    \centering
    \includegraphics[width=0.8\textwidth,trim=0mm 0mm 0mm 0mm]{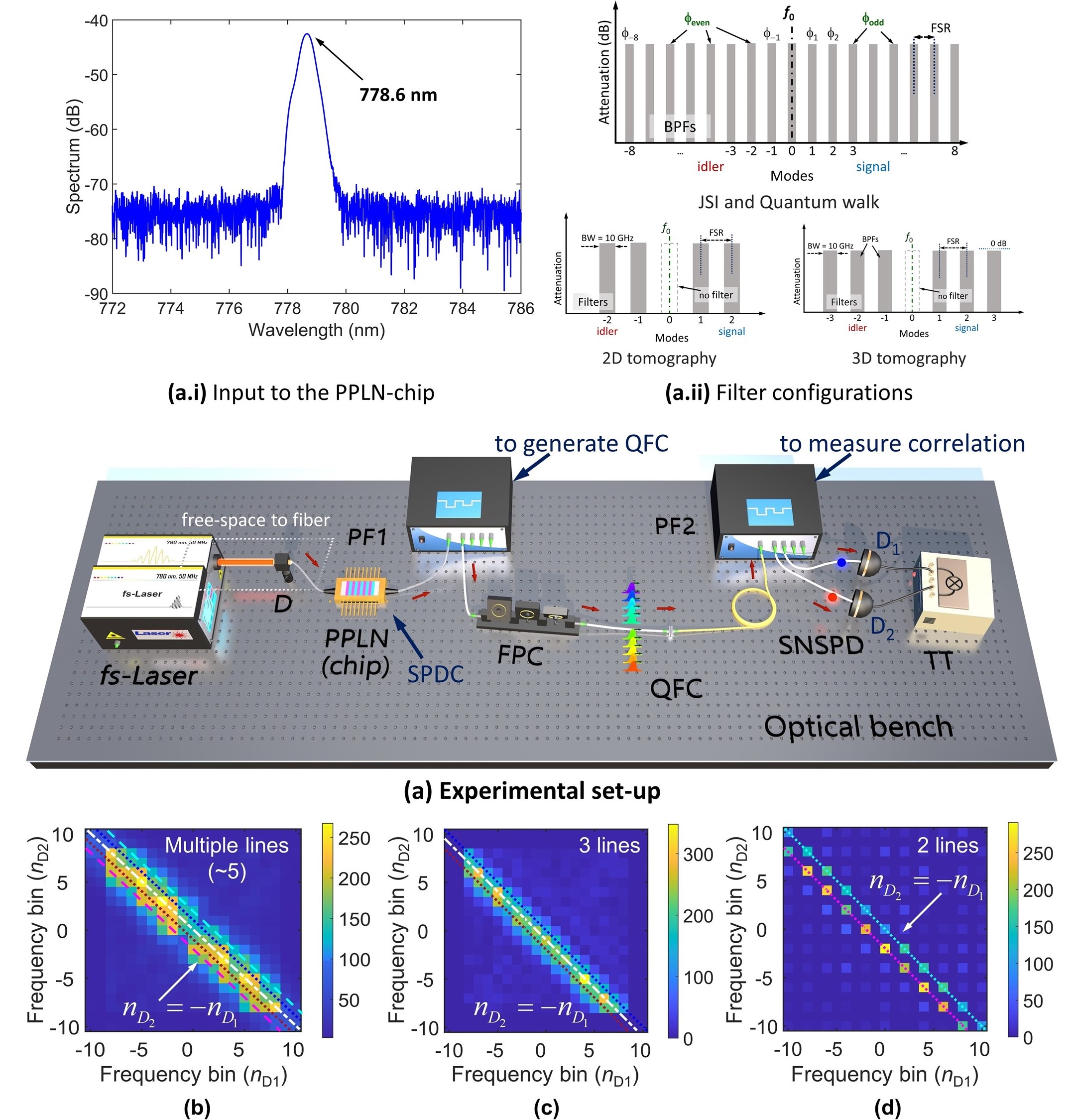}
    \caption{
    \textbf{Demonstration of non-maximally entangled QFCs with variety of JSIs. (a)} Experimental set-up for the generation of QFC in the fs-laser driven PPLN-chip through SPDC. In the insets: \textbf{(a.i)} Spectrally shaped input excitation from the fs-laser to the PPLN-chip,  which is centrally located at about $778.6\pm2$\,nm (measured by an optical spectrometer). \textbf{(a.ii)} Different filter configurations at PF1 to generate a QFC with $17$ frequency-bins, and to perform a quantum walk. The required PF1 configurations to perform quantum state tomography on two (qubits), and three (qutrits) dimensional QFCs are also shown. We can apply a user-defined phase ($\phi$) on each PF1-mask, necessary to generate the projections for the tomography (see Supplementary Materials). JSI of the QFC from \textbf{(b)} \SI{80}{\giga\hertz} bandwidth PPLN-chip, when BPFs at PF1 are placed \SI{25}{\giga\hertz} apart: approximately five anti-diagonal coincidence lines, \textbf{(c)} PPLN with \SI{40}{\giga\hertz} phase-matching bandwidth and same PF1 configuration (\SI{25}{\giga\hertz}), \textbf{(d)} \SI{40}{\giga\hertz} PPLN-chip with \SI{50}{\giga\hertz} GHz PF1 filter-free spectral range (FSR). FPC: fiber polarization controller, SNSPD: superconducting nanowire single-photon detector, D: free-space to fiber, $D_{1, 2}$: detectors, TT: Time tagger. 
    }
    \label{fig: setup JSI}
\end{figure*}

The experimental set-up (Fig: \ref{fig: setup JSI} (a)) used a fs-laser (pulse width $\sim$ \SI{80}{\femto\second}, \SI{50}{\mega\hertz} repetition rate, central wavelength ($\lambda_\mathrm{p} \approx 778.6\,\pm 0.2$\,nm) to excite a PPLN waveguide, generating broadband time-frequency entangled photon pairs with $\sim$ \SI{40}{\nano\meter} bandwidth through SPDC (see the Supplementary). The waveguide output was passed through a programmable filter (PF1) to define the frequency modes of the QFC with $N = 17$ frequency bins of \SI{10}{\giga\hertz} bandwidth each. The output of the PF1 was collected and split by another programmable filter (PF2) to measure photon coincidences. We have two different PPLN chips having overall generation bandwidths of $\sim$ \SI{80}{\giga\hertz} and $\sim$ \SI{40}{\giga\hertz}. Note that the generation bandwidth depends upon the spectral envelope of the pump and the phase-matching bandwidth. Therefore, we can alter the number of antidiagonal lines (and entropy of the system) either by changing the pump bandwidth, phase-matching bandwidth, or by changing the FSR of the PF1, which is demonstrated in Fig. \ref{fig: setup JSI}. As observed in Fig. \ref{fig: setup JSI} (b) and \ref{fig: setup JSI} (c), we obtain about five and three dominant frequency anti-correlation lines from two individual PPLN chips with $\sim$ \SI{80}{\giga\hertz} and $\sim$ \SI{40}{\giga\hertz} generation bandwidths, respectively. BPFs of the PF1 have been placed by \SI{25}{\giga\hertz} apart in both cases. We also attained two anti-diagonal lines in the JSI-diagram for the PPLN-chip with $\sim$ \SI{40}{\giga\hertz} bandwidth when the PF1 filters are set to be \SI{50}{\giga\hertz} apart (Fig. \ref{fig: setup JSI} (d)). Therefore, the proposed excitation and filtering scheme enable the engineering of quantum frequency states from maximally entangled ($D/N \ll 1$) to nearly separable states ($D/N \approx 1$), covering a wide range of entropies. 
%----------------------------------------Steering paper Fig. 3.----------------------------------
%----------------------------------------Steering paper Fig. 3.----------------------------------
%----------------------------------------Steering paper Fig. 4.----------------------------------
\begin{figure*}[t]
    \centering
    \includegraphics[width=0.9\textwidth,trim=0mm 0mm 0mm 0mm]{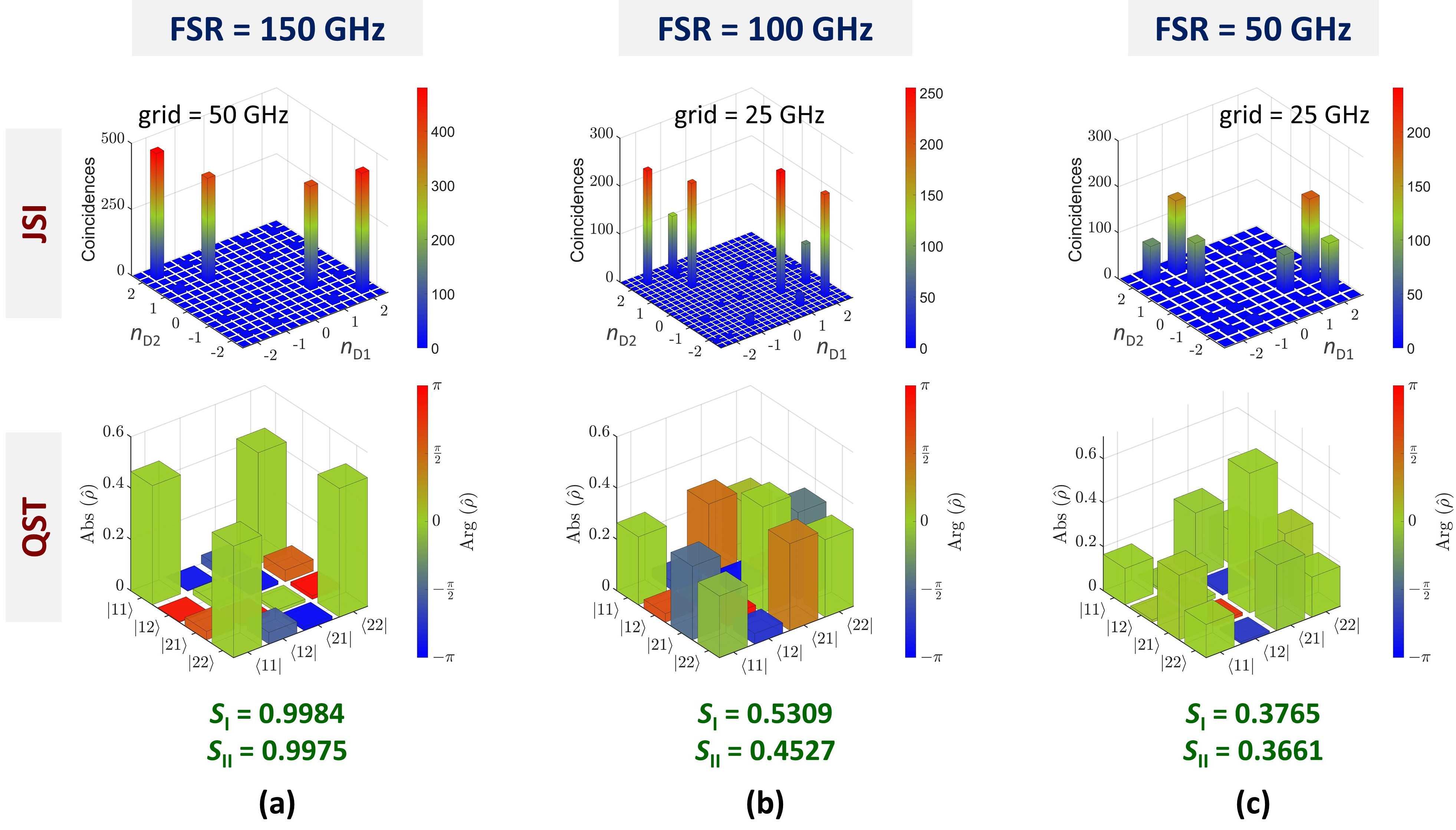}
    \caption{
    \textbf{Quantum state tomography and the entropy calculation for two-dimensional QFC.} JSI (upper row) and the reconstructed density matrices (lower row) of the QFC obtained from $40$ GHz PPLN chip, when the filter separations at PF1 (equivalently the FSR of the QFC) are \textbf{(a)} $150$ GHz, \textbf{(b)} $100$ GHz, \textbf{(c)} $50$ GHz. The filter grid spacings at PF2 while taking the measurements were $50$ GHz, $25$ GHz, and $25$ GHz, for \textbf{(a)}, \textbf{(b)}, and \textbf{(c)}, respectively. For each case, the calculated Von Neumann entropy from the reconstructed density matrix is given. 
    }
    \label{fig: QST 2D}
\end{figure*}
%----------------------------------Steering paper Fig. 4.----------------------------------------
%----------------------------------Steering paper Fig. 5.----------------------------------------
\begin{figure*}[t]
    \centering
    \includegraphics[width=0.6\textwidth,trim=0mm 0mm 0mm 0mm]{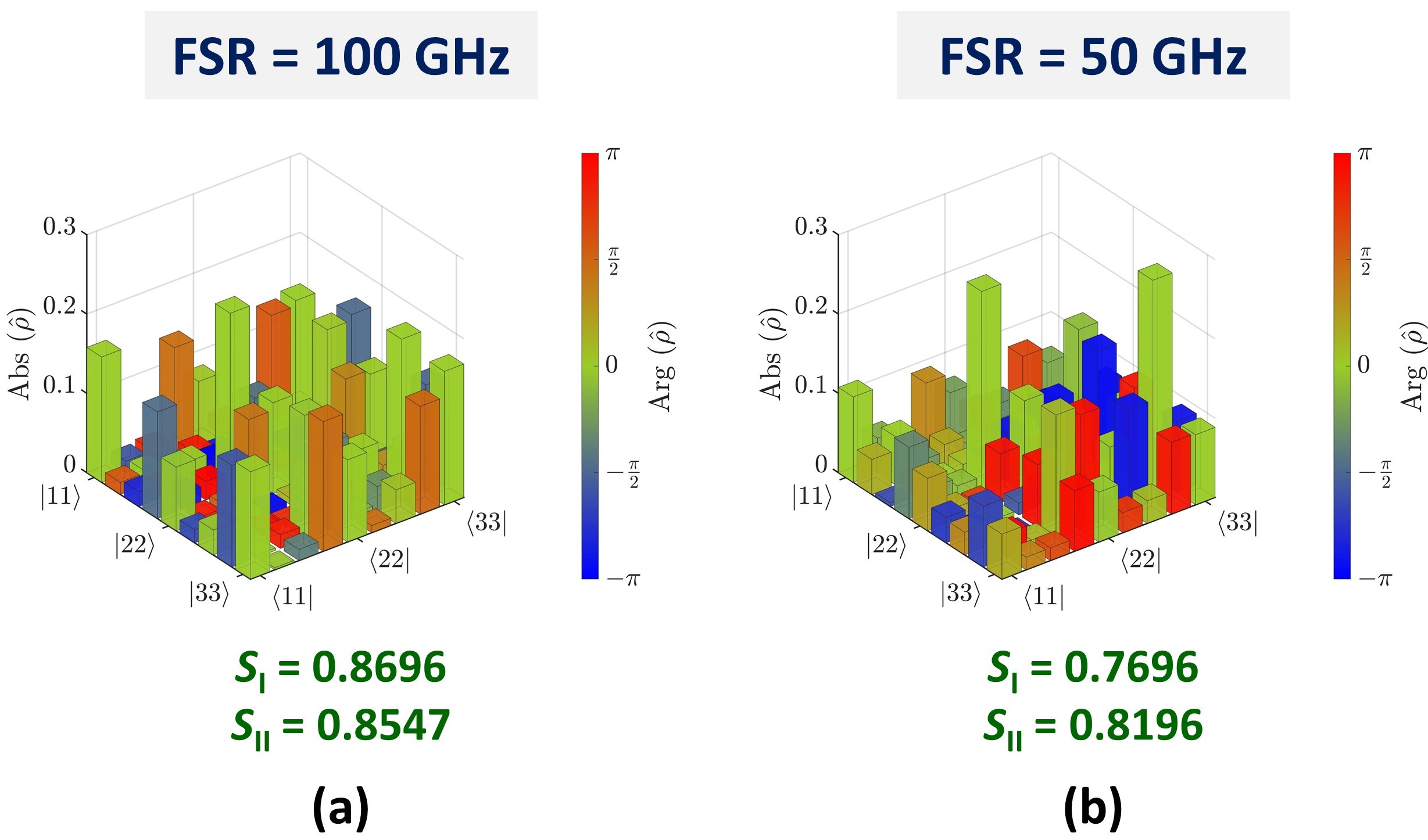}
    \caption{
    \textbf{Quantum state tomography and the entropy calculation of the QFC having three dimensions.} Reconstructed density matrices of the three-dimensional (qutrits) QFC obtained from $40$ GHz PPLN chip, while the filter separations at PF1 (equivalently the FSR of the QFC) are \textbf{(a)} $100$ GHz, and \textbf{(b)} $50$ GHz. For each case, the calculated Von Neumann entropy from the reconstructed density matrix is also listed.  
    }
    \label{fig: QST 3D}
\end{figure*}
%----------------------------------Steering paper Fig. 5.----------------------------------------
\section{Quantum state tomography}
To corroborate our theoretical findings that how the number of frequency anti-correlation lines is related to the entropy, we experimentally reconstructed the density matrix ($\hat{\rho }$) of the QFC through full QST. We performed the QST on the QFCs having dimensions $d = 2$ and $3$, generated from the $\sim$ \SI{40}{\giga\hertz} bandwidth chip. The degeneracy mode, i.e., the central frequency-bin (mode $0$) was omitted by the first programmable filter (PF1) for simplicity. To generate QFCs with different entropies,  the spectral spacing between two adjacent BPFs of the filter array at PF1 was set to \SI{50}{\giga\hertz}, \SI{100}{\giga\hertz}, and \SI{150}{\giga\hertz}, respectively. Note that the filter separations were chosen to be the integer multiple of the phase-modulator’s driving frequency, i.e., \SI{25}{\giga\hertz}. To implement the projectors ($\Psi_\mathrm{proj}$) necessary for QST, an optical phase modulator was used to create sidebands via mode-mixing \cite{kues2017chip}. We calculated the fidelity, purity, and entropy of the QFC from the reconstructed density matrix for each filter setting and compared the values to the ones for the theoretically predicted quantum states. The QFC with multiple antidiagonal lines within the JSI generated through SPDC from a perfectly coherent laser pulse should ideally be a pure quantum state. The prepared quantum state deviated slightly from being a perfectly pure state that corresponds to the predicted QFC as given by Eqs. (\ref{eq S: 2D non-maximal QFC}-\ref{eq S: 3D non-maximal QFC}). We achieved a fidelity close to unity while the overall noise, including phase fluctuations during the measurements, and the spectral incoherence of the pulse-excitation are taken into account. The meticulous details regarding the QST and the formation of the density matrices are provided in the Supplementary information. 

The experiments indicated that  an almost perfect maximally entangled QFC can be produced when the filter separation at the PF1 is \SI{150}{\giga\hertz}. For the QFC with $d = 2$, Fig. \ref{fig: QST 2D} (a), (b), and (c) show the JSIs (upper row) and the corresponding reconstructed density matrices (lower row, the absolute values as bars and the argument color-coded) with filter separations (FSRs) of \SI{150}{\giga\hertz}, \SI{100}{\giga\hertz}, and \SI{50}{\giga\hertz}, respectively. We calculated the reduced density matrices (${{\hat{\rho }}_{\mathrm{I}}}$ and ${{\hat{\rho }}_{\mathrm{II}}}$) by taking the partial traces of the measured density matrices. The Von Neumann entropies $S_\mathrm{I} $ ($S_\mathrm{II}$) obtained from the reduced density matrices ${{\hat{\rho }}_{\mathrm{I}}}$ (${{\hat{\rho }}_{\mathrm{II}}}$) are $0.9984$ ($0.9975$) ($\sim$ maximally entangled state), $0.5309$ ($0.4527$), and $0.3765$ ($0.3661$), respectively. The fidelity (purity) of the measured QFC corresponds to Fig. \ref{fig: QST 2D} (a) (FSR $=$ \SI{150}{\giga\hertz}) with respect to the two-dimensional maximally entangled Bell-state is $95.77\%$ $(85.23\%)$ following the noise model (Supplementary Materials). Note that, for both the cases (b) and (c), non-zero real $\hat{\rho }\mathrm{(2, 1)}$ are the signatures of the co-existing second anti-diagonal QFC line with the central anti-diagonal line. For the FSR $=$ \SI{50}{\giga\hertz}, a greater value of $\hat{\rho }\mathrm{(2, 1)}$  indicates the presence of a brighter second anti-diagonal line, which further reduces the entropy compared to the QFC with a FSR of \SI{100}{\giga\hertz}. The fidelity (purity) attained from the measured density matrices for FSR $=$ \SI{100}{\giga\hertz} and \SI{50}{\giga\hertz} with respect to the theoretically predicted quantum states in the presence of noise are optimized to $95.8\%$ ($85\%$), and $97.48\%$ ($86\%$), respectively. Theoretically predicted Von Neumann entropies of the QFCs with FSR \SI{150}{\giga\hertz}, \SI{100}{\giga\hertz}, and \SI{50}{\giga\hertz} are $1$, $0.489$, $0.361$, respectively, which are indeed in close agreement with our experimentally obtained values.

Thereafter we carried out the full QST to reconstruct the density matrices of the entangled qutrit ($d = 3$) states while the spectral filter separations at PF1 were \SI{100}{\giga\hertz} and \SI{50}{\giga\hertz}. We were restricted by a maximum of $-3$\,dBm input power of the RF-drive. Therefore, it was not possible to perform the QST for the QFC with \SI{150}{\giga\hertz} FSR for the three-dimensional case due to extremely low coincidence counts.  

Fig. \ref{fig: QST 3D} (a) and (b) show the reconstructed density matrices for the QFCs where the filter spacings are \SI{100}{\giga\hertz} and \SI{50}{\giga\hertz}, respectively. The Von Neumann entropies $S_\mathrm{I}$ ($S_\mathrm{II}$) for filter separations \SI{100}{\giga\hertz} and \SI{50}{\giga\hertz} are $0.8696$ ($0.8547$), and $0.7696$ ($0.8196$), respectively. We also calculated other measurements of entanglements such as concurrence associated with each quantum state (irrespective of pure or mixed states) and compared them to the Von Neumann entropy (refer to the Supplementary Materials). All results for different entanglement measures follow a similar trend, i.e., the entanglement increases with the total number of frequency-bins, and decreases with the increase in the number of frequency anti-correlation lines. Such highly adaptive novel qudit states can have potential use in noisy intermediate-scale quantum algorithms, in quantum cryptography, or as a valuable resource for the classification of unknown quantum states using machine learning algorithms.

\section{Steering of quantum walk}
%----------------------------------Steering paper Fig. 6.----------------------------------------
\begin{figure*}[t]
    \centering
    \includegraphics[width=0.9\textwidth,trim=0mm 0mm 0mm 0mm]{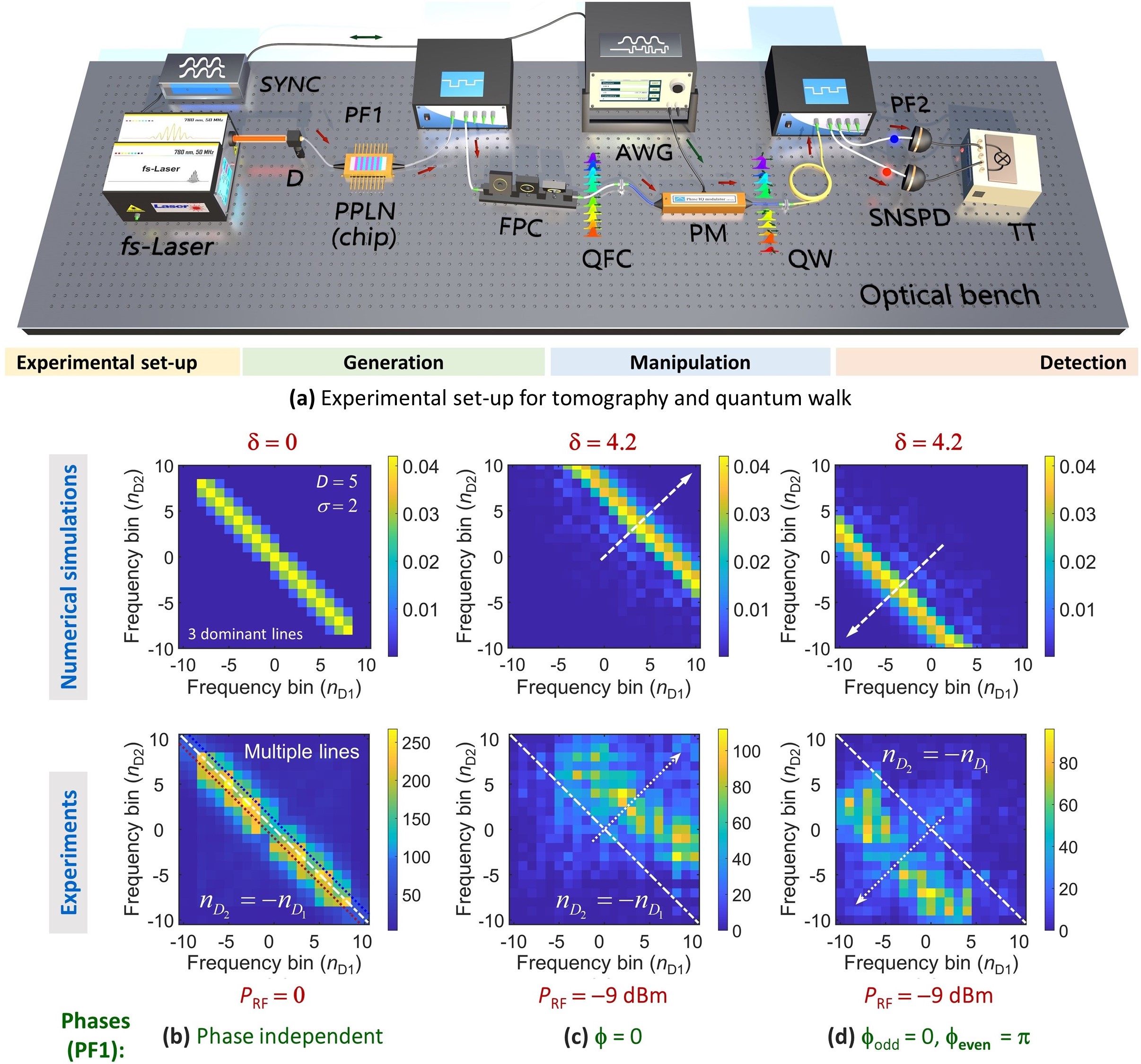}
    \caption{
    \textbf{Steering and coherent control of quantum walk from non-maximally entangled quantum states. (a)} Experimental setup for quantum walk steering. \textbf{(b)} JSI of the QFC without PM: multiple anti-diagonal coincidence lines, \textbf{(c)} QW with PF1-phase $0$, \textbf{(d)} QW with $0$ and $\pi$ alternative phases at PF1. Upper row of the figures \textbf{(b)}--\textbf{(d)}) were obtained from simulations. Notably, to achieve the coherent control of the quantum walk, a synchronization module (SYNC) that synchronizes the modulation frequency of the RF-source that drives the phase modulator (PM) with the laser repetition rate in real-time, is absolutely necessary. PPLN: Periodically poled lithium niobate, PF: Programmable filter, FPC: Fiber polarization controller, AWG: arbitrary waveform generator, SNSPD: superconducting nanowire single-photon detector, TT: Time tagger. 
    }
    \label{fig: QW}
\end{figure*}
%----------------------------------Steering paper Fig. 6.----------------------------------------
A wider variety of continuous QWs was explored based on non-maximally entangled states, i.e., QFC with reduced entropy \cite{omar2006quantum}. We show directional QWs with asymmetric energy transports, where the motion of the QW can be regulated precisely by simply altering the phases of the PF1 (Fig. \ref{fig: QW}). This is initiated from a varying degree of entanglement of a high-dimensional QFC having multiple anti-diagonal lines in the JSI. Such QWs can be used to implement directional graphs or to accelerate search algorithms where the solution space is highly confined in a specific region and can be predicted under certain conditions. 

To achieve the control over QW, we used the same QFC generated within the $\sim$ \SI{40}{\giga\hertz} chip with $N=17$ frequency bins and \SI{25}{\giga\hertz} spacing. A phase-modulator, an arbitrary waveform generator, and a synchronization unit (SYNC) were added to the experimental setup shown in Fig. \ref{fig: setup JSI} (a) to achieve the QW (see Fig. \ref{fig: QW} (a)). As seen from Fig. \ref{fig: QW} (b), due to the broad (\SI{40}{\giga\hertz}) single-mode frequency bandwidth, the JSI had multiple ($5$ with $\sim 3$ dominant) anti-diagonal lines. To simulate such a state, we assume a QFC with $5$ anti-diagonals following a Gaussian distribution with $\sigma = 2$. Such QFC had less entropy than previously generated \cite{imany2020probing} QFCs. The output of the PF1 was connected to a phase-modulator (PM), driven by a sinusoidal RF signal at a frequency that is in congruence via SYNC with the excitation repetition rate, via a synchronization unit (SYNC). The importance of synchronization in phase both in demonstrating directional QW and conducting QST is discussed in the Supplementary Materials. The output of the PM is collected and split by another filter (PF2) to measure photon coincidences by SNSPD, and TT. Fig. \ref{fig: QW} (c) illustrates the QWs when the phases of all frequency modes were set via PF1 to $0$. It is conspicuous that the QW is directional with a tendency to move towards the energetic blue-region. From Fig. \ref{fig: QW} (d), we observe that the walk can be steered towards the red-side if the PF1-phases are alternatively $0$ and $\pi$. 

Obtained results demonstrate all-optical coherent control of the QW direction and entanglement transport via the realization of a non-maximally entangled QFC and adjustment of the phase of the frequency modes. The reported experimental results are in excellent agreement with simulations, and essential for understanding asymmetric QWs and their control for future information processing applications. 

\section{Properties of Quantum walk, coherent control and future scope}
%----------------------------------Steering paper Fig. 7.----------------------------------
\begin{figure*}[t]
    \centering
    \includegraphics[width=0.95\textwidth,trim=0mm 0mm 0mm 0mm]{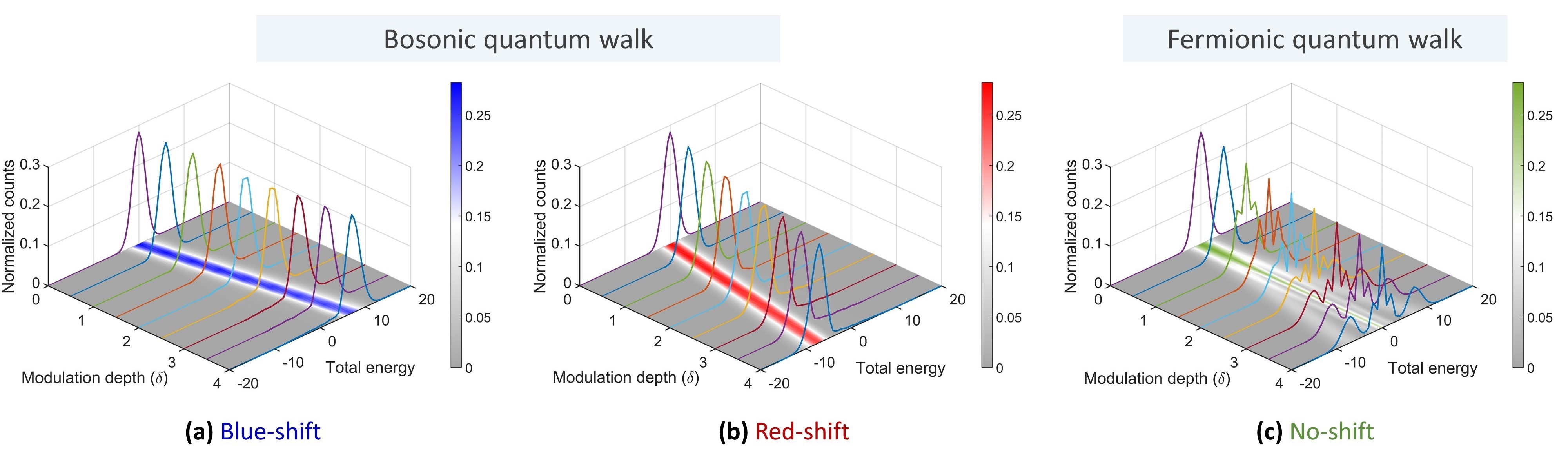}
    \caption{
    \textbf{Energy distribution and transport of the bi-photon QFC with respect to the modulation depth for bosonic, and fermionic QWs.} We observed an overall blue-shift \textbf{(a)}, and red-shift \textbf{(b)} of the total energy possessed by the QFC, in a bosonic configuration, where the phases at PF1 are: $\phi_{\mathrm{odd}}=\phi_{\mathrm{even}}=0$, and $\phi_{\mathrm{odd}}=0, \phi_{\mathrm{even}} =\pi$, respectively. However, for the fermionic QW \textbf{(c)}, which is obtained by a phase-configuration at PF1: $\phi_{\mathrm{odd}}=0, \phi_{\mathrm{even} }=\pi/2$ no energy exchange between the QFC and the RF-field was noticed. 
    }
    \label{fig: QWenergy}
\end{figure*}
%----------------------------------------Steering paper Fig. 7.----------------------------------
%----------------------------------------Steering paper Fig. 8.----------------------------------
\begin{figure*}[t]
    \centering
    \includegraphics[width=0.95\textwidth,trim=0mm 0mm 0mm 0mm]{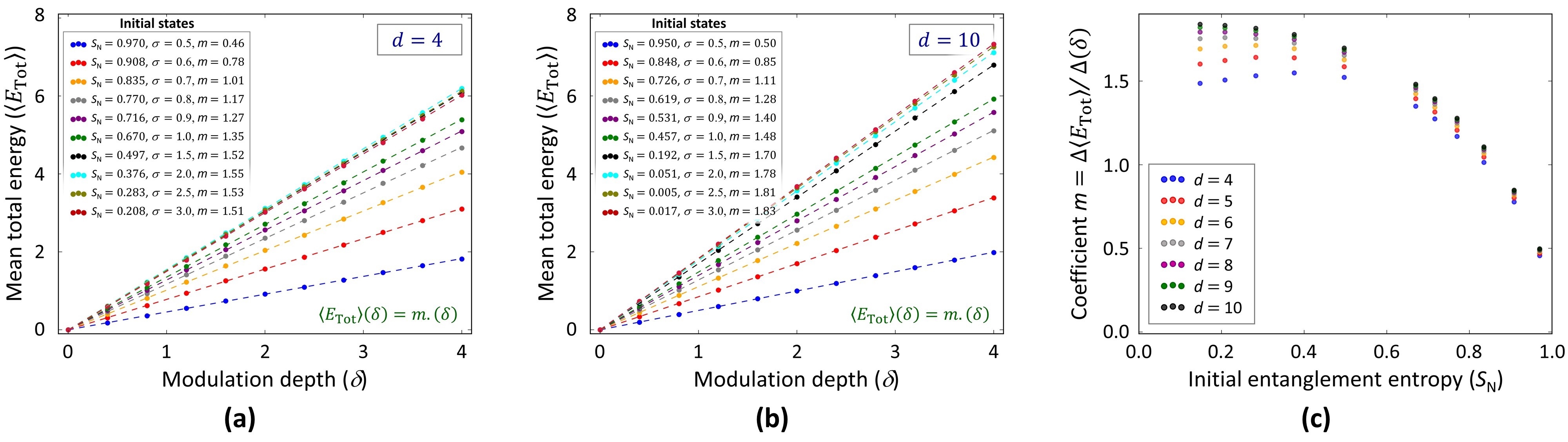}
    \caption{
    \textbf{Dependencies of the mean of the total energy ($E_{\mathrm{Tot}}$) on the modulation depth ($\delta$), initial normalized entanglement entropy ($S_{\mathrm{N}}$), and the dimension ($d$) of the QFC. (a)} Mean value of $E_{\mathrm{Tot}}$ with respect to modulation depth for $d = 4$, \textbf{(b)} Mean value of $E_{\mathrm{Tot}}$ with respect to modulation depth for $d = 10$, \textbf{(c)} Slope ($m$) of the change in acquired energy by the QFC from the EOM with respect to the initial normalized entanglement entropy ($S_{\mathrm{N}}$) for the QFCs having different dimensions ($4$ to $10$).    
    }
    \label{fig: QWdependency}
\end{figure*}
%----------------------------------Steering paper Fig. 8.----------------------------------
We further investigate that how the bosonic and fermionic QW \cite{imany2020probing} can be controlled through different parameters such as the dimension and entropy of the initial quantum state. Fig. \ref{fig: QWenergy} shows the change in total energy of a QFC with respect to the modulation depth for the bosonic and fermionic quantum walk. A blue (red) shift in total energy of the QFC indicates an energy exchange from (to) the RF-modulator to (from) the QFC, whereas in fermionic QW, no such energy transfer was noted. In the Supplementary Materials, we provide a detailed theoretical analysis explaining the origin of such propensities in the movement of the quantum walks. We mathematically established (see Supplementary Materials) that such energy transfer (gain or loss) between the quantum system and the modulator can only happen in QWs commencing from non-maximally entangled states. Moreover, we emphasize that the directional property of the QW, here propelled by the mutual interaction of the RF-modulator and the QFC, being a direct consequence of reduced entropy, is a fundamental phenomenon irrespective of the system it was realized. Therefore, the steering phenomenon of QW is equally valid for other reported quantum states with reduced entropies \cite{haldar2021multi, u2006generation} or with versatile JSI \cite{kumar2014controlling}.       

Fig. \ref{fig: QWdependency} presents a quantitative analysis to estimate the energy transfer from the RF-field to the QFC with respect to the modulation depth, dimension of the comb, and initial normalized entropy for the case of bosonic QW (i.e., $\phi_\mathrm{odd} = \phi_\mathrm{even} = 0$). From Fig. \ref{fig: QWdependency} (a) and (b) we conclude that the energy transfer increases with an increase in dimension. From Fig. \ref{fig: QWdependency} (c) it is explicitly visible that the energy transfer to a QFC from the RF-field increases when the initial entropy of the QFC decreases.    

Finally, we explored that such directional QW initiated from a non-maximally entangled state excited by a fs-laser, unlike the QW excited by a CW-laser as demonstrated in \cite{imany2020probing}, cannot be obtained without using a synchronized unit (see Supplementary Materials)

\section{Conclusions and future directions}
In conclusion, we demonstrated high-dimensional two-photon entangled states with versatile joint spectral intensities having multiple frequency anti-correlation lines. We performed quantum state tomography and reconstructed the density matrices of the quantum frequency combs with near unit fidelities through which we verified the maiden generation of high-dimensional quantum photonic states with tunable entropies varying from $0.35$ to $1$.  Such non-maximally entangled QFCs are crucial resources for certifying several QIP protocols such as teleportation, cloning, deletion, entanglement concentration, distillation, purification, error correction, and have the potential to play pivotal roles in the classification of unknown quantum states through machine learning or in NISQ technologies. 

More importantly, we demonstrated the steering of quantum walk by launching such high-dimensional two-photon quantum states with varying degrees of entanglement. For the first time, we presented an in-depth quantitative analysis of the dependency of bosonic and fermionic QW on the dimensions and the initial entropies of the QFCs spawning the same QW. It is worthwhile to note that in addition to the effortless modification of the QW-depth without altering the device footprint as reported by Imany \textit{et al}.\cite{imany2020probing}, unlike Bose-Einstein condensate \cite{weiss2015steering}, here we accomplish an unprecedented command  over the transport properties of two photons in a bosonic tight binding model with remarkable ease. We anticipate at least two different directions for future research. First, our research may benefit the investigation of few-photon dynamics in tight-binding models with non-trivial topological band structure (e.g., Su-Schrieffers Heeger model in $1$D) by non-Hermitian electro-optic modulation. Recently, such investigations have been reported for classical light \cite{leefmans2022topological, wang2021generating}. By using quantum light, new dynamical features such as topological phases and localization to emerge at the single-photon level can be expected, which can pave the way to topologically protected photonic quantum states. Second, the QW dynamics presented in this article can be enriched by means of a hybrid coined-continuous QW, which simultaneously uses the time- and frequency degree of freedom and thereby combines the features of coined quantum walks with bosonic tight-binding dynamics. Generation of hyperentangled photons in both frequency and time as well as their coherent manipulation have already been demonstrated experimentally \cite{ reimer2019high}. Hybrid coined-continuous quantum walks may readily find applications in quantum search algorithms or new variants of boson sampling, and quantum information transport. We also envisage such QWs due to their biased nature of energy transport can consequently be implemented in forecasting the financial status of the market, explaining critical physical phenomena, e.g., photosynthesis, or the source of cognizance in the human brain by emulating them in real-time \cite{wergen2011record, codling2008random}. 
%===================================================
\let\oldaddcontentsline\addcontentsline% Store \addcontentsline
\renewcommand{\addcontentsline}[3]{}% Make \addcontentsline a no-op
\let\addcontentsline\oldaddcontentsline% Restore \addcontentsline

\noindent{}\textbf{Acknowledgements.---}
The project was supported by the German federal ministry of education and research, Quantum Futur Program (PQuMAL). R. H. acknowledges the financial support provided by Alexander von Humboldt Stiftung to conduct the research. R. H. also acknowledges A. Dutt (Stanford Univ.), S. Chatterjee (Raman Research Institute), U. Sen, and A. K. Pati (Harish-Chandra Research Institute), C. M. Chandrashekhar (IISC Bangalore), J. Capmany (Polytechnic Univ. of Valencia), Eleni Diamanti (Sorbonne University, Paris Centre for Quantum Computing), and A. Prabhakar (IIT Madras) for valuable insights.\\

\noindent{}\textbf{Author contributions.---}
R. H., and M. K. conceived the work. R. H., P. R., T. B., performed the theoretical analysis and the numerical simulations. A. K. K., R. H., and M. K. designed and assembled the devices, and characterized the fabricated PPLN-chip. R. J. and R. H. performed the quantum state tomography. R. H. performed the quantum walk measurements and prepared the first draft. Finally, R. H., P. R., R. J., S. B., N. T. J., and M. K. analyzed the results. M. K. supervised the project. All authors discussed the results and edited the manuscript.\\

%\noindent{}
%\textbf{Competing interests.---}

%\textbf{Methods.---}

\clearpage
\onecolumngrid

% \usepackage{graphicx}
% \usepackage{amsmath}
% \usepackage{amssymb,mathtools}
% \usepackage{epstopdf}
% \usepackage{braket}
% \usepackage{color}
% \usepackage[normalem]{ulem}
% \usepackage[dvipsnames]{xcolor}
% \DeclarePairedDelimiter\floor{\lfloor}{\rfloor}

% \DeclareGraphicsExtensions{.eps}

% \newcommand{\comment}[1]{\textit{\small\textcolor{red}{#1}}} 
% \newcommand*{\rom}[1]{\expandafter\@slowromancap\romannumeral #1@}
% \newcommand{\red}[1]{\textcolor{red}{#1}} 
% \newcommand{\blue}[1]{\textcolor{blue}{#1}} 
% \newcommand{\green}[1]{\textcolor{OliveGreen}{#1}} 
% \newcommand{\bl}[1]{\textcolor{blue}{#1}} 

% \newcommand{\violet}[1]{\textcolor{violet}{#1}} 
% \newcommand{\op}[1]{\widehat{#1}}
% \newcommand{\ad}{\hat{a}}
% \newcommand{\au}{\hat{a}^\dagger}
% \newcommand{\bd}{\hat{b}}
% \newcommand{\bu}{\hat{b}^\dagger}

% %%%%%%%%%%%%%%%%%%%%%%%%%%%%%%%%%%%%%%%%
% %%%%%%%%%%%%%%%%%%%%%%%%%%%%%%%%%%%%%%%%
% \begin{document}

\begin{center}
\textbf{\large Supplementary Materials: Steering of Quantum Walks through the Coherent Control of High-dimensional Bi-photon Quantum Frequency Comb with Tunable State Entropies}\\[.5cm]
  
Raktim Haldar,$^{1, 2, 3, *}$ Robert Johanning,$^{1, 2, 3}$ Philip R\"ubeling,$^{1, 2, 3}$ Anahita Khodadad Kashi,$^{1, 2, 3}$ Thomas B{\ae}kkegaard,$^{1, 2, 4}$ Surajit Bose,$^{1, 2, 3}$ Nikolaj Thomas Zinner,$^{4, 5}$ and Michael Kues$^{1, 2, 3, \dagger}$
\\[.1cm]
{\itshape 
$^1$Institute of Photonics, Leibniz University Hannover, Nienburger Stra\ss e 17, 30167 Hannover, Germany

$^2$Hannover Centre for Optical Technologies, Leibniz University Hannover, Nienburger Stra\ss e 17,\\ 30167 Hannover, Germany

$^3$Cluster of Excellence PhoenixD (Photonic, Optics, and Engineering – Innovation Across Disciplines), \\Leibniz University Hannover, Hannover, Germany

$^4$Kvantify APS, DK-2300, Copenhagen S, Denmark\\

$^5$Department of Physics and Astronomoy, Aarhus University, Aarhus C DK-8000, Denmark\\ }
\end{center}

\setcounter{equation}{0}
\setcounter{figure}{0}
\setcounter{table}{0}
\setcounter{page}{1}
\makeatletter
\renewcommand{\theequation}{S.\arabic{equation}}
\renewcommand{\thefigure}{S.\arabic{figure}}
In this document, we describe the details of the theoretical modeling and experimental analysis of the generation of bi-photon quantum states with tunable entropies and steering of quantum walk (QW) from non-maximally entangled quantum frequency combs (QFCs). In section \ref{Subsection: Multiple JSIs}, at first we discussed various methods to generate a QFC with multiple anti-correlation lines in the joint spectral intensity (JSI). Thereafter we derive the mathematical expressions for such non-maximally entangled QFCs having multiple anti-correlation lines in the JSI. Calculations of both the absolute and normalized entropies of the QFCs are also explained in great detail. In the next section (section \ref{Subsection: Simulation of QW}) we present the mathematical modeling and the numerical simulations of the QW from non-maximally entangled QFCs. Characterization and emission bandwidth from the periodically poled lithium niobate (PPLN) chip are discussed in section \ref{Subsection: PPLN}. In section \ref{Subsection: EOM}, we focus on the method of quantum state tomography (QST). We present a comprehensive details on setting up the electro-optic modulator (EOM) for QST, Bell-test measurements, fidelity calculations and other relevant information. Finally, we culminate the Supplementary Information by investigating the significance of using a synchronization unit (SYNC) to obtain the directional quantum walk in section \ref{Subsection: SYNC}.

%\tableofcontents

\clearpage
\section{Theory and Simulation of the QFC with Multiple JSI lines}
\label{Subsection: Multiple JSIs}

\noindent{}\textbf{Generation.} In this section, we discuss the theory, simulation, and experimental schemes to generate QFC with multiple anticorrelation lines in the JSI. In our previous works, we demonstrated that there could be two different methods to generate the QFC with multiple anticorrelation lines in the JSI, having tunable entropies: (i) Exciting a nonlinear waveguide (here, PPLN) with a pulsed-laser having a finite and significantly broad bandwidth \cite{haldar2021high_sup}, (ii) multi-chromatic continuous-wave (CW) excitations to several resonant modes of a nonlinear microresonator (MR) \cite{haldar2021multi_sup}. Although in this work we adopted the first approach with an expense of reduced brightness, the latter approach is robust, does not necessitate post-selection, and produces well-defined anticorrelation lines with tunable entropies, which can be theoretically predicted precisely, giving rise to versatile JSIs within a much smaller footprint, compared to coupled resonators \cite{kumar2014controlling_sup}. Here, we preferred the first approach because of the unavailability of suitable bandpass filters covering the desired resonant modes of the MR, and our freedom to choose the desired FSR of the QFC, essential for performing the tomography.   
%----------------------------------Steering paper Fig. S.1.----------------------------------
\begin{figure*}[htbp]
    \centering
    \includegraphics[width=0.85\textwidth,trim=0mm 0mm 0mm 0mm]{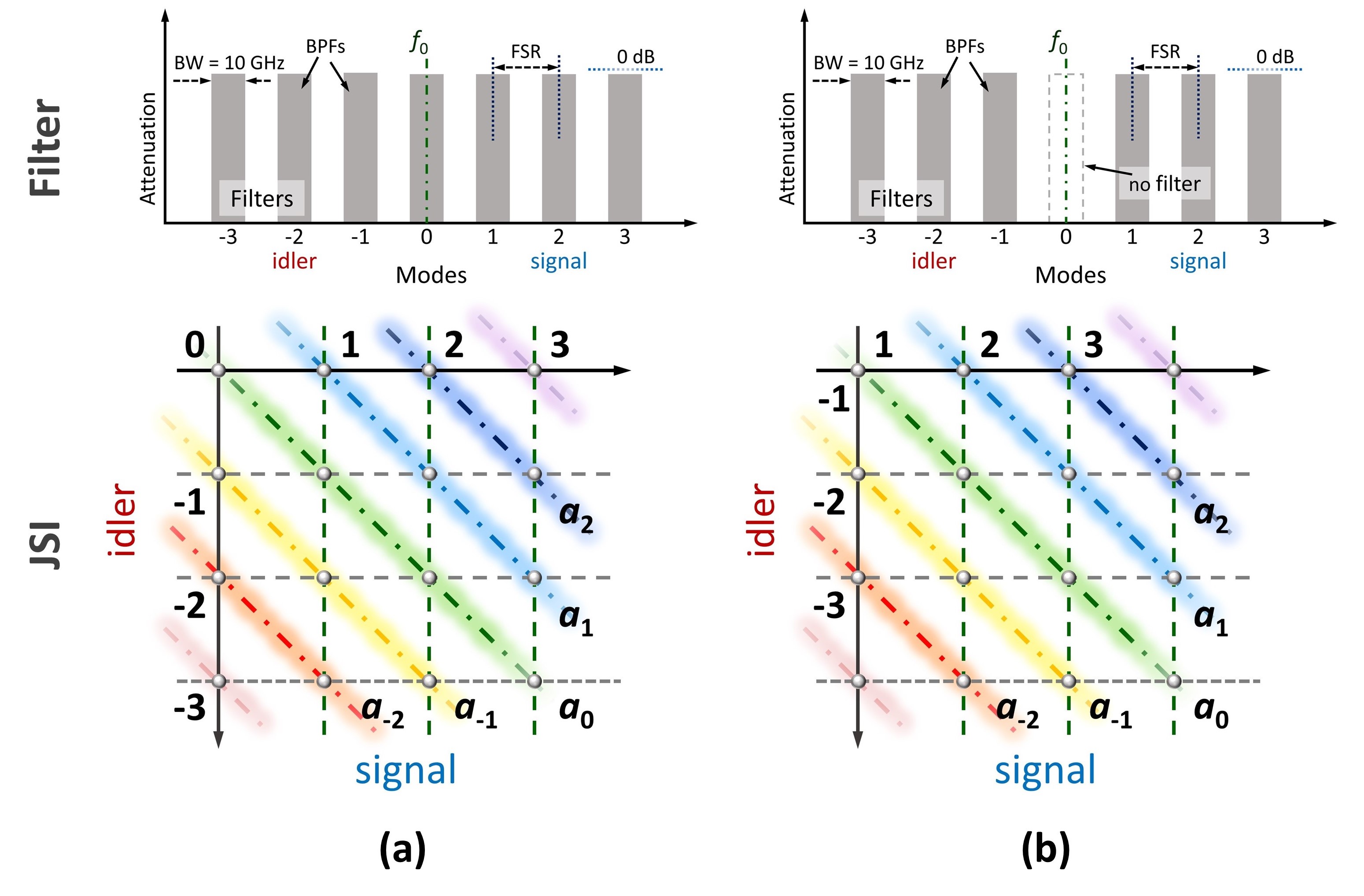}
    \caption{
    \textbf{Scheme to generate a QFC in a PPLN waveguide with the help of a programmable filter.} Filter configuration at PF1 and the corresponding JSI when the degenerate resonant mode of the QFC is \textbf{(a)} present, and \textbf{(b)} removed.}
    \label{fig S: JSI 0th mode}
\end{figure*}
%--------------------------------Steering of QW paper Fig. S.1.----------------------------------

We derived the mathematical equations for the QFCs having multiple anti-diagonal lines in the JSI. Expression for the $q$-th anti-diagonal ($\left| {{\psi }_{q}} \right\rangle $) of the QFC as shown in Fig. \ref{fig S: JSI 0th mode} (a) is given by, 
%------------------------------------------------------------------------------------------------
\begin{align}
\left| {{\psi }_{q}} \right\rangle ={{\left| {{\psi }_{q}} \right\rangle }^{\left( a \right)}}=\frac{1}{\sqrt{{\left( N-1 \right)}/{2}\;-\left( \left| q \right|-1 \right)}}\sum\limits_{k=\left| q \right|}^{{(N-1)}/{2}\;}{\left\{ \begin{array}{*{35}{l}}
   {{\left| k+q \right\rangle }_{s}}{{\left| -k \right\rangle }_{i}}, & \text{for the lower diagonals }(q<0)  \\
   {{\left| k \right\rangle }_{s}}{{\left| -k \right\rangle }_{i}}, & \text{for the central diagonal }(q=0)  \\
   {{\left| k \right\rangle }_{s}}{{\left| -\left( k-q \right) \right\rangle }_{i}}, & \text{for the upper diagonals }(q>0)  \\
\end{array} \right.}
\label{eq S: QFC1}
\end{align}
%------------------------------------------------------------------------------------------------
If the number of diagonals present in the JSI is $D$, then the final expression of the QFC can be written as,
%------------------------------------------------------------------------------------------------
\begin{align}
\left| {{\psi }_{\mathrm{QFC}}} \right\rangle =\frac{\sum\limits_{q=1}^{D}{{{a}_{q}}\left| {{\psi }_{q}} \right\rangle }}{\sum\limits_{q=1}^{D}{{{\left| {{a}_{q}} \right|}^{2}}}},
\label{eq S: aq dist}
\end{align}
%------------------------------------------------------------------------------------------------
The distribution of ${{\left| {{a}_{\mathrm{q}}} \right|}^{2}}$ that plays a crucial role in determining the entropy of the QFC depends upon the phase-matching condition of the PPLN and the spectral shape of the input excitation. If the FSR of the QFC is kept very small compared to the overall generation bandwidth of the QFC (JSI bandwidth), the distribution of $a_\mathrm{q}$ can be assumed flat over a limited number of diagonals. Otherwise, it is reasonable to assume that $a_\mathrm{q}$ follows a Gaussian distribution with standard deviation $\sigma$, and mean $\mu$, which is given by 
%------------------------------------------------------------------------------------------------
\begin{align}
{{a}_{q}}\left( q \right)={{a}_{0}}{{e}^{-{{\left( \frac{\left| q \right|-\mu }{\sigma } \right)}^{2}}}}
\label{eq S: aq Gaussian}
\end{align}
%------------------------------------------------------------------------------------------------
Here, $\mu$ accounts for the asymmetry in phase-matching and pump excitation, whereas, $\sigma$ takes care of the generation bandwidth. In most of our simulations, we assume $a_\mathrm{0}=1$, and $\mu=0$ (symmetric nonlinear conversion) for simplicity. Note that, without a loss of generality, we can omit the central degenerated mode of the QFC using PF1 for simplicity, as shown in Fig. \ref{fig S: JSI 0th mode} (b). In that case, the entropy of the QFC slightly changes with respect to the reported one and the rank $r$ of the reduced density matrix coincides with the dimension ($d = (N-1)/2$) of the QFC. Similarly, the expression for QFC will be
%------------------------------------------------------------------------------------------------
\begin{align}
\left| {{\psi }_{q}} \right\rangle ={{\left| {{\psi }_{q}} \right\rangle }^{\left( b \right)}}=\frac{1}{\sqrt{{\left( N-1 \right)}/{2}\;-\left| q \right|}}\sum\limits_{k=\left| q \right|+1}^{{(N-1)}/{2}\;}{\left\{ \begin{array}{*{35}{l}}
   {{\left| k+q \right\rangle }_{s}}{{\left| -k \right\rangle }_{i}}, & \text{for the lower diagonals }(q<0)  \\
   {{\left| k \right\rangle }_{s}}\left| -k \right\rangle {}_{i}, & \text{for the central diagonal }(q=0)  \\
   {{\left| k \right\rangle }_{s}}{{\left| -\left( k-q \right) \right\rangle }_{i}}, & \text{for the upper diagonals }(q>0)  \\
\end{array} \right.}
\label{eq S: QFC2}
\end{align}
%------------------------------------------------------------------------------------------------
Alternative, the QFC is given by Eq. (\ref{eq S: QFC2}) can also be described in terms of dimension ($d$) of the comb within a single equation as
%------------------------------------------------------------------------------------------------
\begin{align}
\left| {{\psi }_{\mathrm{QFC}}} \right\rangle ={{\sum\limits_{r=-p}^{p}{\sum\limits_{l=1+\left| r \right|}^{d-\left| r \right|}{\left( \left( 1+2\left| p \right| \right)\left( d-2\left| r \right| \right) \right)}}}^{-\frac{1}{2}}}{{\left| r+l \right\rangle }_{s}}{{\left| r-l \right\rangle }_{i}}
\label{eq S: QFC3v1Philip}
\end{align}
%------------------------------------------------------------------------------------------------
Where the total number of diagonals is $D = 2p+1$. Considering a Gaussian distribution Eq. (\ref{eq S: QFC3v1Philip}) can be rewritten within a single equation as,
%------------------------------------------------------------------------------------------------
\begin{align}
\left| {{\psi }_{\mathrm{QFC}}}\left( {{\sigma }'} \right) \right\rangle =Z\left( {{\sigma }'} \right){{\sum\limits_{r=-p}^{p}{\sum\limits_{l=1+\left| r \right|}^{d-\left| r \right|}{\left( d-2\left| r \right| \right)}}}^{-\frac{1}{2}}}{{e}^{-\frac{{{r}^{2}}}{2{{{{\sigma }'}}^{2}}}}}{{\left| r+l \right\rangle }_{s}}{{\left| r-l \right\rangle }_{i}}
\label{eq S: QFC3v2Philip}
\end{align}
%------------------------------------------------------------------------------------------------
Where $\sigma'$ is the width of the Gaussian envelope and the $Z\left( {{\sigma }'} \right)$ is the normalization coefficient that has to be determined numerically. Note that, at the limit of ${\sigma }'\to \infty $  the entropy of the QFC given by Eq. (\ref{eq S: QFC3v2Philip}) reduced to $0$, whereas, for ${\sigma }'\to 0$   the QFC becomes maximally entangled \cite{kues2017chip_sup}. Subscripts `$s$' and `$i$' with the bra-ket notations are used to represent the signal and the idler, respectively.\\
%----------------------------------Steering paper Fig. S.2.----------------------------------
\begin{figure*}[htbp]
    \centering
    \includegraphics[width=0.85\textwidth,trim=0mm 0mm 0mm 0mm]{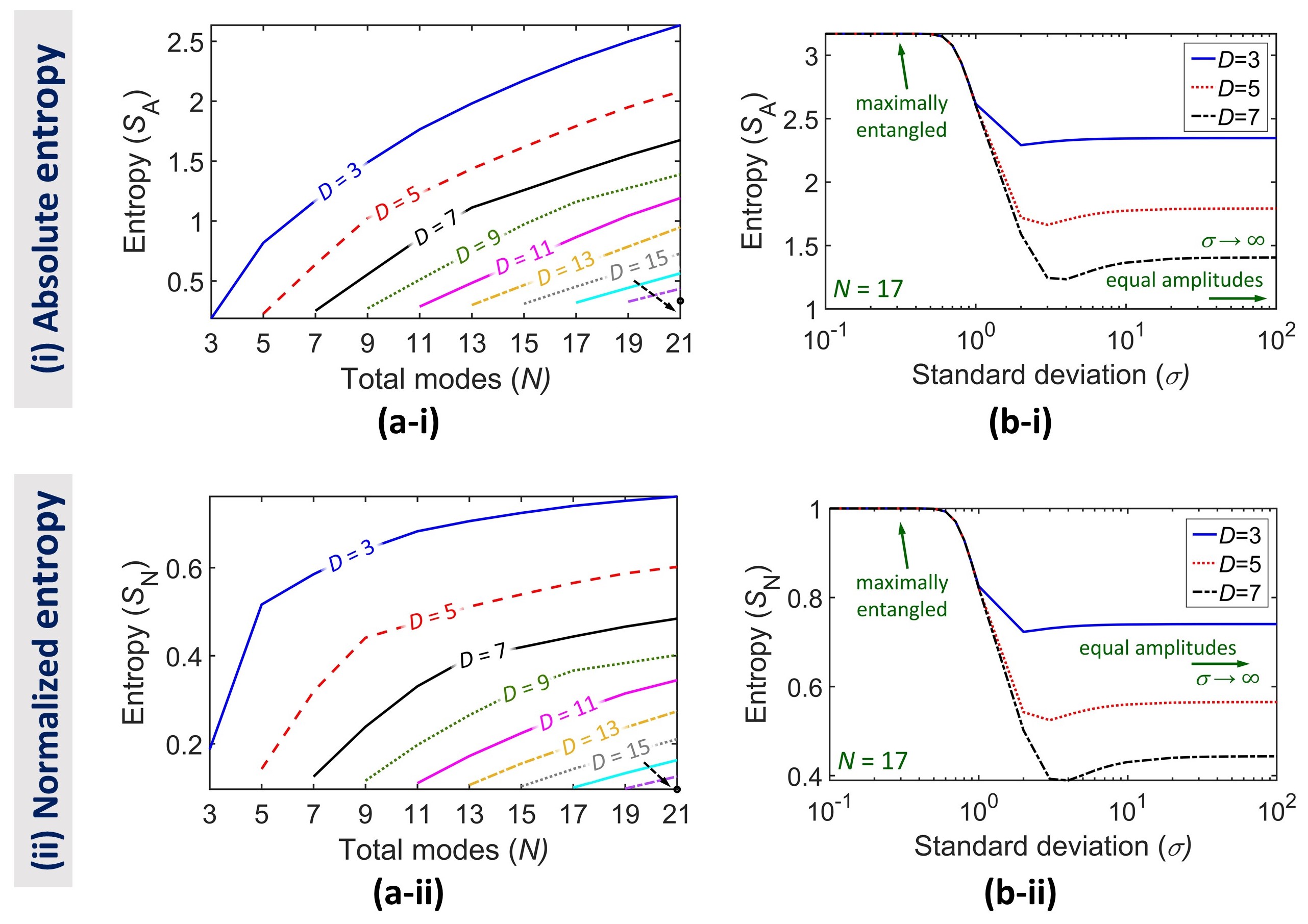}
    \caption{
    \textbf{Calculated absolute ($S_{\mathrm{A}}$) and normalized ($S_{\mathrm{N}}$) entropies. (a-i)} Entropy $S_{\mathrm{A}}$, and \textbf{(a-ii)} $S_{\mathrm{N}}$ as a function of the total number of QFC-modes $N$ with varying JSI-diagonals $D$s in the JSI when all the diagonals are equiprobable. \textbf{(b-i)} $S_{\mathrm{A}}$, and \textbf{(b-ii)} $S_{\mathrm{N}}$ with the standard deviation ($\sigma$) when the JSI follows the normal distribution. BW: bandwidth.}
    \label{fig S: Absolute vs normalized entropy}
\end{figure*}
%--------------------------------Steering of QW paper Fig. S.2.----------------------------------
%----------------------------------Steering paper Fig. S.3.----------------------------------
\begin{figure*}[htbp]
    \centering
    \includegraphics[width=0.85\textwidth,trim=0mm 0mm 0mm 0mm]{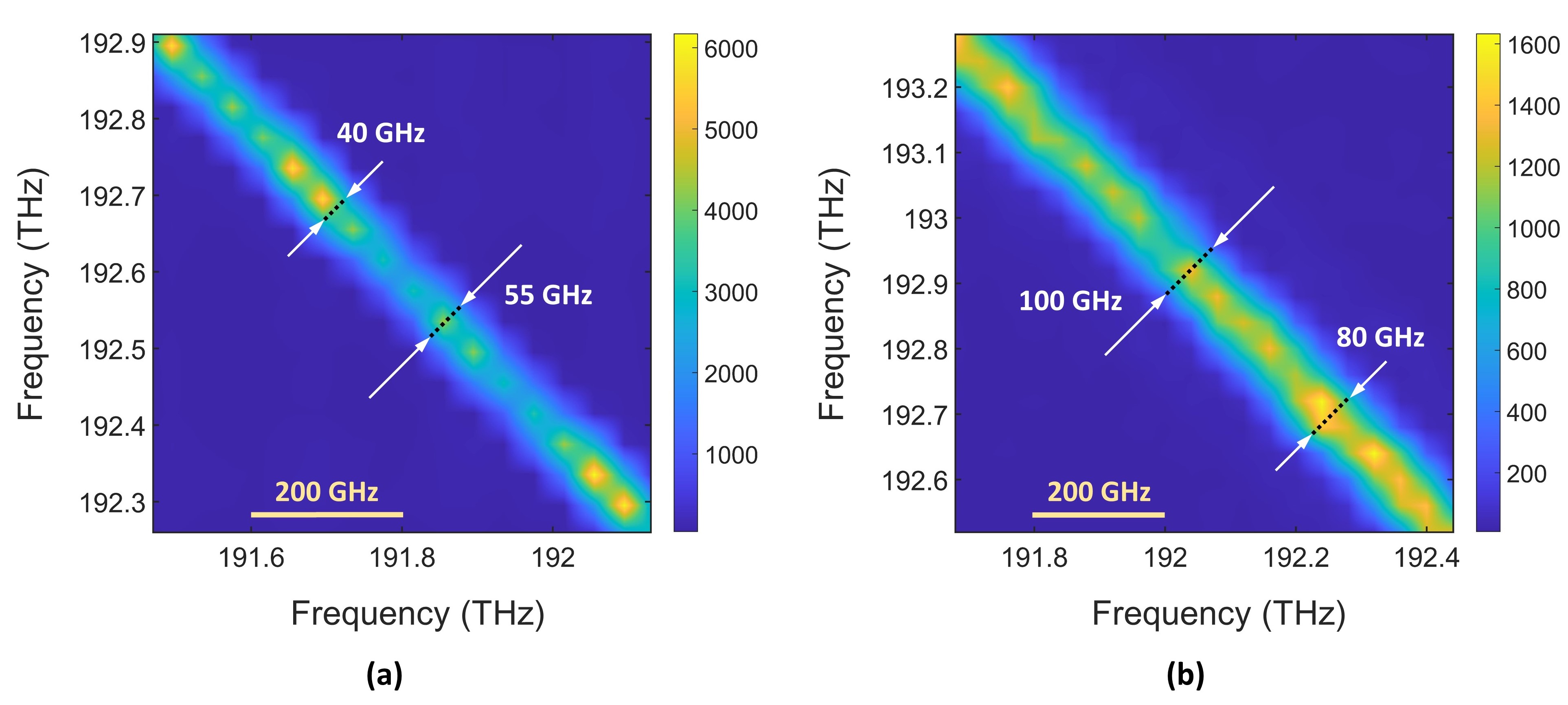}
    \caption{
    \textbf{Measured joint spectral intensities (JSIs) from the two PPLN waveguides.} JSI for \textbf{(a)} PPLN waveguide from Covesion Ltd. with $\sim$\,\SI{40}{\giga\hertz} bandwidth, \textbf{(b)} PPLN waveguide designed by AdvR Inc., having almost double, i.e., $\sim$\,\SI{80}{\giga\hertz} bandwidth. The input power measured by a power meter at the input of both the waveguides was \SI{6}{\micro\watt}. The acquisition time was \SI{16000}{\second} and \SI{12000}{\second} for \textbf{(a)} and \textbf{(b)}, respectively.}
    \label{fig S: PPLN JSIs}
\end{figure*}
%--------------------------------Steering of QW paper Fig. S.3.----------------------------------

\noindent{}\textbf{Calculation of the entropy.} Von Neumann entropy of the biphoton QFC can be considered as a measurement of the entanglement between the two parties, viz., signal and idler, being in the superpositions of several resonant frequency modes. Therefore, to compute the entropy, at first, we determined the density matrix corresponds to the QFC given by any of the Eqs. (\ref{eq S: QFC1}--\ref{eq S: QFC3v2Philip}). Then we take the partial trace to calculate the reduced density matrix $\hat{\rho }$. By diagonalizing the reduced density matrix $\hat{\rho }$  of rank $r$, we computed the absolute von Neumann entropy ${{S}_{\mathrm{A}}}=-\sum{{{\lambda }_{\mathrm{m}}}}{{\log }_{2}}\left( {{\lambda }_{\mathrm{m}}} \right)$, where $\lambda_{\mathrm{m}}$ are the eigenvalues of $\hat{\rho }$. One can calculate the dimension of the QFC by evaluating the Schmidt number from the density matrix. Note that the absolute entropy always increases with the dimension of the quantum state. Therefore, to compare the dependency of the entropy on various parameters as shown in Fig. \ref{fig S: Absolute vs normalized entropy} such as the total number of diagonals in the JSIs of QFCs having different dimensions, we introduced the normalized entropy ${{S}_{\mathrm{N}}}=-\sum{{{\lambda }_{\mathrm{m}}}}{{\log }_{r}}\left( {{\lambda }_{\mathrm{m}}} \right)$. Here instead of $2$, the rank $r$ of the reduced density matrix is used as the logarithmic base. The rank of the reduced density matrix is synonymous with the dimension ($d$) of the QFC shown in Fig. \ref{eq S: QFC1} (b) where the degenerated mode is absent. $S_{\mathrm{N}}$ can vary from $0$ to $1$ for the completely separable state to the maximally entangled state, respectively. At the end of this Supplementary, we also discussed the validity of entanglement entropy in relation to the obtained experimental data from QST and discussed other techniques to quantify entanglement.

\section{Theory and Simulation of the Quantum Walk}
\label{Subsection: Simulation of QW}

\noindent{}\textbf{Mathematical modeling and numerical simulation.} An electro-optic phase modulator (EOM) enables coherent mixing of the frequency modes of a QFC, and is thereby an indispensable component for the manipulation of quantum photonic states exploiting the frequency degree of freedom of photons. An EOM is also crucial to realize a conventional SU ($2$), or a high-dimensional SU ($d$) beam splitter in frequency domains. Effectively, an EOM can transfer a photon in $m$-th resonant mode to ($m+n$) mode with a probability amplitude of $J_\mathrm{n}(\delta)$, where the argument of the Bessel function of the first kind $\delta$ is the modulation depth related to the RF-power that drives the EOM. Therefore, in terms of the Unitary operator \cite{imany2020probing_sup}, the action of the EOM can be written as
%------------------------------------------------------------------------------------------------
\begin{align}
U=\sum\limits_{n=-\infty }^{\infty }{{{J}_{\mathrm{n}}}(\delta ){{a}_{\mathrm{m+n}}}^{\dagger }{{a}_{\mathrm{m}}}}
\label{eq S: QW unitary}
\end{align}
%------------------------------------------------------------------------------------------------
Where ${{a}_{\mathrm{m}}}\left( a_{m}^{\dagger } \right)$ is the annihilation (creation) operator related to frequency mode $m$. The unitary operator acting any of the quantum frequency states $\left| {{\psi }_{\mathrm{QFC}}} \right\rangle $   given by Eq. (\ref{eq S: QFC1}--\ref{eq S: QFC3v2Philip}) yields the required quantum walk corresponding to that initial quantum state. The QW can also be simulated from the evolution of the Hamiltonian of the QFC in Heisenberg’s picture, considering the Hamiltonians of the EOM \cite{hu2020realization_sup} given by Eq. (\ref{eq S: QW Hamiltonian}). Nevertheless, this method is time-consuming and numerically demanding. It is interesting that the similarity between the Hamiltonian of the EOM and the tight-binding Hamiltonian opens plethora of possibilities towards realizing quantum walk with topologically protected states.
%------------------------------------------------------------------------------------------------
\begin{align}
H=\sum\limits_{m}{\frac{\Omega }{2}\left( {{a}_{m}}^{\dagger }{{a}_{m+1}}+\text{h}\text{. c}\text{.} \right)}
\label{eq S: QW Hamiltonian}
\end{align}
%------------------------------------------------------------------------------------------------
Where ${{\left| \Omega  \right|}^{2}}$ denotes the hopping probability of the photons between the adjacent frequency modes.\\

\noindent{}\textbf{Mathematical explanation of the steering.} This section aims to explain why for maximally entangled states (MES) no steering is observed, whereas, for the partially entangled state (PES), there is the steering of the quantum walk. It is also important to note that steering is related to an energy transfer from the QFC to the bosonic mode of the RF modulator or vice versa. If $H_{\mathrm{RF}}$ and $H_{\mathrm{QFC}}$ denote the Hamiltonians related to the RF-field and the QFC, from energy conservation, we can write: $\dot{H}_{\mathrm{RF}} = - \dot{H}_{\mathrm{QFC}}$. Therefore, the expression for the total energy of the QFC is given by,
%------------------------------------------------------------------------------------------------
\begin{align}
{{H}_{\text{QFC}}}=\sum\limits_{\alpha }{\hbar }{{\omega }_{\alpha }}b_{\alpha }^{\dagger }{{b}_{\alpha }},
\label{eq S: QFC Hamiltonian}
\end{align}
%------------------------------------------------------------------------------------------------
where ${{\omega }_{\alpha }}={{\omega }_{\mathrm{CEO}}}+\alpha {{\omega }_{\mathrm{RF}}}$ are equidistant frequencies of the comb. $\omega_{\mathrm{CEO}}$ and $\omega_{\alpha}$ are the carrier-envelope offset (CEO) and RF modulation angular frequencies, respectively. The time evolution of the total energy in the Heisenberg picture combined with the semiclassical approach of electro-optic modulation \cite{ hu2020realization_sup, capmany2010quantum_sup, rueda2019resonant_sup} is given by,
%------------------------------------------------------------------------------------------------
\begin{align}
{{\dot{H}}_{\text{QFC}}}=\frac{d}{dt}(\sum\limits_{\alpha }{\hbar }{{\omega }_{\alpha }}b_{\alpha }^{\dagger }{{b}_{\alpha }})=\sum\limits_{\alpha }{\hbar }{{\omega }_{\alpha }}(\dot{b}_{\alpha }^{\dagger }{{b}_{\alpha }}+b_{\alpha }^{\dagger }{{\dot{b}}_{\alpha }})
\label{eq S: QFC time evolution Hamiltonian}
\end{align}
%------------------------------------------------------------------------------------------------
According to Heisenberg’s equation of motion, the time evolution of the annihilation operator $b_\alpha$ is determined by the interaction Hamiltonian and thus,
%------------------------------------------------------------------------------------------------
\begin{align}
{{\dot{b}}_{\alpha }}=\frac{i}{\hbar }[\sum\limits_{\beta }{{{g}_{1}}}b_{\beta +1}^{\dagger }{{b}_{\beta }}+g_{1}^{*}b_{\beta }^{\dagger }{{b}_{\beta +1}},\text{ }{{b}_{\alpha }}]=\frac{i}{\hbar }({{g}_{1}}\sum\limits_{\beta }{[}b_{\beta +1}^{\dagger }{{b}_{\beta }},\text{ }{{b}_{\alpha }}]+g_{1}^{*}\sum\limits_{\beta }{[}b_{\beta }^{\dagger }{{b}_{\beta +1}},\text{ }{{b}_{\alpha }}])
\label{eq S: QFC time evolution b alpha}
\end{align}
%------------------------------------------------------------------------------------------------
Using bosonic commutation relations $[b_{\beta}, b_{\alpha}^{\dagger}] = \delta_{\beta\alpha}$ and explicitly calculating both commutators, $[b_{\beta+1}^{\dagger}b_{\beta}, b_{\alpha}] = - \delta_{\alpha\beta+1} b_{\beta}$ and $[b_{\beta}^{\dagger}b_{\beta+1}, b_{\alpha}] = - \delta_{\alpha\beta} b_{\beta+1}$, one finds 
%------------------------------------------------------------------------------------------------
\begin{align}
{{\dot{b}}_{\alpha }}=-\frac{i}{\hbar }({{g}_{1}}{{b}_{\alpha -1}}+g_{1}^{*}{{b}_{\alpha +1}})
\label{eq S: QFC time evolution b alpha v2}
\end{align}
%------------------------------------------------------------------------------------------------
%------------------------------------------------------------------------------------------------
\begin{align}
\dot{b}_{\alpha }^{\dagger }=+\frac{i}{\hbar }(g_{1}^{*}b_{\alpha -1}^{\dagger }+{{g}_{1}}b_{\alpha +1}^{\dagger })
\label{eq S: QFC time evolution b alpha dagger v2}
\end{align}
%------------------------------------------------------------------------------------------------
Next, inserting the time derivatives into equation Eq. (\ref{eq S: QFC time evolution b alpha}) and grouping terms with ${{g}_{1}}$, $g_{1}^{*}$ together one obtains,
%------------------------------------------------------------------------------------------------
\begin{align}
   {{{\dot{H}}}_{\text{QFC}}} & ={{g}_{1}}\sum\limits_{\alpha }{i}{{\omega }_{\alpha }}(b_{\alpha +1}^{\dagger }{{b}_{\alpha }}-b_{\alpha }^{\dagger }{{b}_{\alpha -1}})+g_{1}^{*}\sum\limits_{\alpha }{i}{{\omega }_{\alpha }}(b_{\alpha -1}^{\dagger }{{b}_{\alpha }}-b_{\alpha }^{\dagger }{{b}_{\alpha +1}}) \nonumber \\
 & =i{{g}_{1}}\sum\limits_{\alpha }{({{\omega }_{\alpha }}-{{\omega }_{\alpha +1}})}\text{ }b_{\alpha +1}^{\dagger }{{b}_{\alpha }}+ig_{1}^{*}\sum\limits_{\alpha }{({{\omega }_{\alpha +1}}-{{\omega }_{\alpha }})}\text{ }b_{\alpha }^{\dagger }{{b}_{\alpha +1}} \nonumber \\
 & =-i{{g}_{1}}{{\Omega }_{\text{FSR}}}\sum\limits_{\alpha }{b_{\alpha +1}^{\dagger }}{{b}_{\alpha }}+ig_{1}^{*}{{\Omega }_{\text{FSR}}}\sum\limits_{\alpha }{b_{\alpha }^{\dagger }}{{b}_{\alpha +1}}  
\label{eq S: QFC time evolution Hamiltonian v2}
\end{align}
%------------------------------------------------------------------------------------------------
 where $\left( {{\omega }_{\alpha +1}}-{{\omega }_{\alpha }} \right)={{\Omega }_{\mathrm{FSR}}}$ (alternatively the modulation frequency) is used. For further discussions, it's useful to use the following symbols for simplicity: $\chi =\sum\nolimits_{\alpha }{b_{\alpha +1}^{\dagger }{{b}_{\alpha }}}$ and its Hermitian conjugate: ${{\chi }^{\dagger }}=\sum\nolimits_{\alpha }{b_{\alpha }^{\dagger }{{b}_{\alpha +1}}}$. Hence, Eq. (\ref{eq S: QFC time evolution Hamiltonian v2}) can be given by, $\dot{H}_{\mathrm{QFC}} = -i g_{1} \Omega_{\mathrm{FSR}} \chi + i g_{1}^{*} \Omega_{\mathrm{FSR}} \chi^{\dagger}$. 
 
 Now, to explain why no net energy transfer or steering can be observed for the MES, consider the high-dimensional maximally entangled state,
 %------------------------------------------------------------------------------------------------
\begin{align}
\left| {{\Psi }^{(\text{MES})}} \right\rangle =\frac{1}{\sqrt{d}}\sum\limits_{l}{\text{ }\left| l,-l \right\rangle }
\label{eq S: MES}
\end{align}
%------------------------------------------------------------------------------------------------
Note that, $\langle \Psi^{\mathrm{MES}} | \chi | \Psi^{\mathrm{MES}} \rangle = \langle \Psi^{\mathrm{MES}} | \chi^{\dagger} | \Psi^{\mathrm{MES}} \rangle = 0$, as scattering a single photon in the signal or idler branch onto a neighboring mode, breaks the symmetry of the maximally entangled state. Thereby it has no overlap with the original state and vanishes. Hence, for the expectation value of the energy transfer one evaluates,
%------------------------------------------------------------------------------------------------
\begin{align}
   \text{tr}({{{\dot{H}}}_{\mathrm{QFC}}}{{\rho }^{(\text{MES})}}) ^ = -i{{g}_{1}}{{\Omega }_{\text{FSR}}}\langle {{\Psi }^{\text{MES}}}|\chi |{{\Psi }^{\text{MES}}}\rangle +ig_{1}^{*}{{\Omega }_{\text{FSR}}}\langle {{\Psi }^{\text{MES}}}|{{\chi }^{\dagger }}|{{\Psi }^{\text{MES}}}\rangle = 0 
\label{eq S: MES E-transfer}
\end{align}
%------------------------------------------------------------------------------------------------
which coincides with the results observed in the experiments. It should be emphasized that the lack of steering is not a unique property of the maximally entangled state (consider the state $\left| 1,-1 \right\rangle $ for example) and therefore cannot be used as an entanglement witness. 

In order to explain the steering and energy transfer for PES consider the simplest example which includes phases of $\phi = \{0, \pi \}$ applied to both frequency modes, i.e.,
%------------------------------------------------------------------------------------------------
\begin{align}
\left| {{\Psi }^{(\text{PES})}} \right\rangle =\frac{1}{2}\left( \left| 1,-1 \right\rangle +{{e}^{i\phi }}\left| 1,-2 \right\rangle +{{e}^{i\phi }}\left| 2,-1 \right\rangle +\left| 2,2 \right\rangle  \right) 
\label{eq S: PES example}
\end{align}
%------------------------------------------------------------------------------------------------
A short calculation reveals that $\langle \Psi^{\mathrm{PES}} | \chi | \Psi^{\mathrm{PES}} \rangle = \langle \Psi^{\mathrm{PES}} | \chi^{\dagger} | \Psi^{\mathrm{PES}} \rangle = \pm 1$ (positive sign for $\phi = 0$ and negative sign for $\phi = \pi$). Hence, for the total energy transfer this results in,
%------------------------------------------------------------------------------------------------
\begin{align}
  \text{tr}({{{\dot{H}}}_{\mathrm{QFC}}}{{\rho }^{(\text{PES})}}) & = -i{{g}_{1}}{{\Omega }_{\text{FSR}}}\langle {{\Psi }^{\text{PES}}}|\chi |{{\Psi }^{\text{PES}}}\rangle +ig_{1}^{*}{{\Omega }_{\text{FSR}}}\langle {{\Psi }^{\text{PES}}}|{{\chi }^{\dagger }}|{{\Psi }^{\text{PES}}}\rangle \nonumber \\
 & =-i{{g}_{1}}{{\Omega }_{\mathrm{FSR}}}+ig_{1}^{*}{{\Omega }_{\mathrm{FSR}}}=-i|{{g}_{1}}|{{\Omega }_{\mathrm{FSR}}}{{e}^{i{{\phi }_{\mathrm{RF}}}}}+i|{{g}_{1}}|{{\Omega }_{\mathrm{FSR}}}{{e}^{-i{{\phi }_{\mathrm{RF}}}}} \nonumber \\ 
 & = 2|g_{1}| \Omega_{\mathrm{FSR}} \sin(\phi_{\mathrm{RF}})
\label{eq S: PES E-transfer}
\end{align}
%------------------------------------------------------------------------------------------------
Here, it is assumed that the laser and the RF-modulator are synchronized with each other, the RF phase is stabilized to be ${{\phi }_{\mathrm{RF}}}=\pi /2$. For the energy transfer, it then follows that,
%------------------------------------------------------------------------------------------------
\begin{align}
\text{tr}({{\dot{H}}_{\mathrm{QFC}}}{{\rho }^{(\text{PES})}})=\text{ }\left\{\begin{matrix}
   +2|{{g}_{1}}|{{\Omega }_{\mathrm{FSR}}},\text{ for }\phi =0  \\
   -2|{{g}_{1}}|{{\Omega }_{\mathrm{FSR}}},\text{ for }\phi =\pi, \\
\end{matrix}\right.
\label{eq S: PES E-transfer values}
\end{align}
%------------------------------------------------------------------------------------------------
Hence, for uniform phases a red shifted frequency steering is observed, while for alternating phases a blue shifted frequency steering is present. Therefore, the conclusions from the mathematical calculations strongly agree with both with the numerical simulations of the quantum walk and the experimental results we obtained. 

\section{Characterization of the PPLN chip}
\label{Subsection: PPLN}

\noindent{}\textbf{PPLN chip and the experimental setup.} We have two standard fiber-coupled PPLN waveguides, one from Covesion Ltd. and another from AdvR Inc. The lithium niobite waveguides were doped with MgO and fabricated through proton exchange method, which allow a very broad transparent window. The chip from Covesion contains a \SI{2.5}{\centi \meter} long $5\%$-MgO doped type-$0$ PPLN waveguide with a single frequency mode bandwidth of $\sim\,40$--$55$\,GHz. Another chip from AdvR Inc. has a \SI{40}{\milli\meter} long type-$0$ PPLN waveguide with a period of $\sim$\,\SI{17}{\micro\meter} and a single frequency mode bandwidth of $\sim$\,$80$--$100$\,GHz. Typical JSIs from the Covesion and the AdvR are shown in the left (a) and the right (b) panels of the Fig. \ref{fig S: PPLN JSIs}.\\

%----------------------------------Steering paper Fig. S.4.----------------------------------
\begin{figure*}[htbp]
    \centering
    \includegraphics[width=0.95\textwidth,trim=0mm 0mm 0mm 0mm]{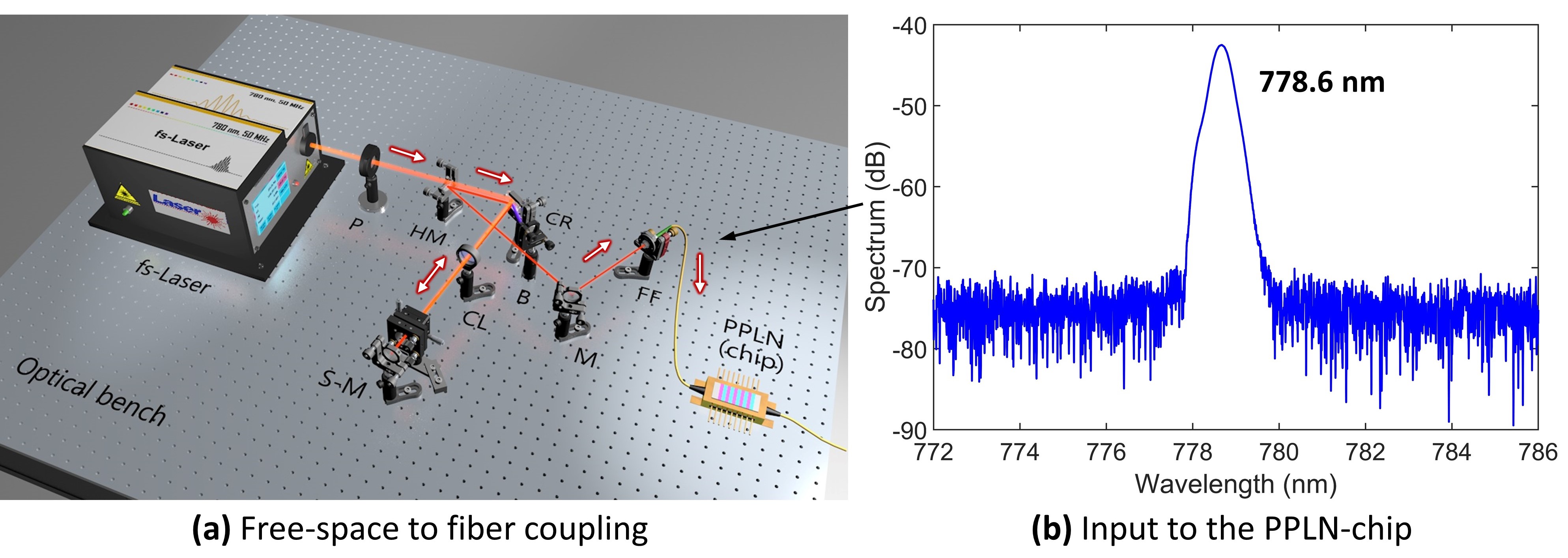}
    \caption{
    \textbf{Preparing the input excitation to the PPLN-chip. (a)} Light coupling and spectral shaping setup from the fs-laser to the PPLN-chip, \textbf{(b)} Typical input spectrum to the PPLN-chip.}
    \label{fig S: PPLN input setup}
\end{figure*}
%--------------------------------Steering of QW paper Fig. S.4.----------------------------------

\noindent{}\textbf{Free-space to fiber coupling setup.} At first, the output of the fs-laser (C-fiber $780$, Menlo systems, operating power \SI{50}{\milli\watt} at \SI{780}{\nano\meter}, \SI{50}{\mega\hertz} repetition rate, \SI{80}{\femto\second} pulse-width) was passed through a polarizer (P), over a half-mirror (HM), and was incident on a diffracting crystal (CR) (Fig. \ref{fig S: PPLN input setup} (a)). The output of the CR was spectrally shaped by an adjustable slit connected to a mirror at its back-end (S-M), after being guided by a collimator (CL). Whereas, the higher-order outputs from the CR were blocked by a blocker (B). The output of the S-M is slightly tilted so that it can be reflected via the HM to another mirror (M), which finally directed the beam towards the free-space to fiber (FF) coupler, connected at the input port of our PPLN-chip. The spectral profile of the input to the PPLN chip is shown in Fig. \ref{fig S: PPLN input setup} (b). Throughout the experiment (tomography and quantum walk), we maintained the input power of the PPLN waveguide approximately at \SI{175}{\micro\watt}.

\section{Quantum state tomography and electro-optic phase modulator}
\label{Subsection: EOM}

\noindent{}\textbf{Synchronization of the EOM to the laser repetition rate.} The EOM was driven by a single sinusoidal radio-frequency (RF) tone at \SI{25}{\giga\hertz}. It was used for the coherent mixing of the frequency modes by creating sidebands, which were spectrally separated by the integral multiples of the EOM driving frequency. The amplitude of the $n$-th sideband at a particular RF power can be described by the $n$-th order Bessel function of the first kind with an argument related to the RF power. For the two-dimensional QSTs, we mixed two adjacent modes with a selected FSR of \SI{50}{\giga\hertz} (\SI{100}{\giga\hertz}, \SI{150}{\giga\hertz}) coherently in the center between the modes, so we used the $1$st ($2$nd, $3$rd) sideband created by the EOM by applying RF power of $-16$\,dBm ($-12$\,dBm, $-9.4$\,dBm). In order to mix three adjacent frequency modes with a selected FSR of \SI{50}{\giga\hertz} (\SI{100}{\giga\hertz}) at the central position with equal intensities, two sidebands, namely $0$th and $2$nd ($0$th and $4$th), must have equal intensities, which is the case with an EOM driving power of $-16$\,dBm ($-9$\,dBm).

The repetition rate of the pulsed laser is about \SI{50}{\giga\hertz} and it has to be perfectly stabilized with respect to the EOM driving frequency to enable coherent measurements. This is done by an interferometer in the synchronization unit viz. SYNC (RRE-Syncro Repetition Rate Stabilizer by MenloSystems), which compares the EOM frequency and the laser repetition rate real-time, and dynamically modifies the cavity length through an electrically controlled stage inside the laser, thereby rectifying the laser repetition rate. Without this synchronization, the resulting states are not pure but incoherent mixtures of the included frequency components.\\

\noindent{}\textbf{EOM offset compensation.} The EOM adds an additional phase to the sidebands which depends on the power and the frequency of the RF-source and scales proportionally with the number of the sideband. We measure this phase between two resonance modes by creating the superposition state of both signal and both idler modes and rotating the phase of only one mode in the first programmable filter (PF1). Thereby the measured coincidence counts followed a sinusoidal curve, where the counts were maximized when the phase of the PF1 was compensated by the phase of the EOM. We permanently set this phase to PF1 to further conduct the rest of the experiments.\\

\noindent{}\textbf{Detailed experimental setup for the quantum state tomography.} For the QST, we need to manipulate the quantum states and mix them in such a way that quantum interference can occur. Therefore, we sent the photon pairs to PF1,  where an attenuation and phase mask is applied to the QFC to create the desired projections. This allows any user-defined phase to be added to each frequency mode. A bandwidth of \SI{20}{\giga\hertz} was chosen for the transmitted frequency bins. Then the photons pass through an EOM, where they are coherently mixed.

At last, PF2 collected the photons with a bandwidth of \SI{25}{\giga\hertz} and divided them into two parts, viz. signal and idler. The PF2 was connected to SNSPDs followed by a time tagger for coincidence measurement with a resolution of \SI{5}{\pico\second} \cite{kues2017chip_sup}. Due to the finite laser bandwidth, the timing jitter in the detectors and the dispersion of the fibers, the time difference between the detection of the two photons of a coincidence count is distributed in a histogram following a Lorentz curve. We selected those counts that laid in a symmetric window with size of \SI{105}{\pico\second} around the center, which is more than the FWHM. We were able to select this broad window, as the noise level is only of the order of $0.05$ counts per \SI{5}{\pico\second}.\\

\noindent{}\textbf{Theoretical description of the generated QFC and the Fidelity calculations.} For a conventional QFC with equiprobable frequency modes that has only one anticorrelation line in the JSI \cite{kues2017chip_sup} within the observed frequency range, ideally, we expect to measure a Bell-state, which can be given by,
%------------------------------------------------------------------------------------------------
\begin{align}
\left| \psi  \right\rangle =\sum\limits_{k=1}^{d}{{{c}_{k}}{{\left| k \right\rangle }_{s}}{{\left| k \right\rangle }_{i}}=\frac{1}{\sqrt{d}}}\sum\limits_{k=1}^{d}{{{\left| k \right\rangle }_{s}}{{\left| k \right\rangle }_{i}}}={{\left| \psi  \right\rangle }_{\mathrm{Bell}}}\text{ with }\sum\limits_{k=1}^{d}{{{c}_{k}}=1}
\label{eq S: High-dimensional Bell-state}
\end{align}
%------------------------------------------------------------------------------------------------
As we eventually reduced the FSR at PF1, multiple anti-diagonal correlation lines in the JSI gradually appeared, which also contributed to the observed quantum states, (see eq. (\ref{eq S: QFC1})). For the two-dimensional states, we expect the state to be of the form,
%------------------------------------------------------------------------------------------------
\begin{align}
\left| \psi  \right\rangle =\frac{1}{\sqrt{1+{{c}_{q=1}}^{2}}}\left( {{\left| \psi  \right\rangle }_{\mathrm{Bell}}}+{{e}^{i\pi \theta }}{{c}_{q=1}}\left| 21 \right\rangle  \right)=\frac{1}{\sqrt{1+{{c}_{q=1}}^{2}}}\left( \frac{\left| 11 \right\rangle +\left| 22 \right\rangle }{\sqrt{2}}+{{e}^{i\pi \theta }}{{c}_{q=1}}\left| 21 \right\rangle  \right)
\label{eq S: 2D non-maximal QFC}
\end{align}
%------------------------------------------------------------------------------------------------
Where $c_{q=1}$ is the amplitude factor for the JSI line with $q=1$, i.e. the first upper JSI line. For the two-dimensional subsystem of the QFC, only the state $\left| 21 \right\rangle$ is contained in that line. The asymmetry, i.e. the absence of the lower JSI line with $q=-1$, is due to an asymmetry in the pump profile relative to the PPLN split point, set by the slit (S-M), as shown in Fig. \ref{fig S: PPLN input setup}.
For the tomography of the three-dimensional states, the expected state is of the form given by,
%------------------------------------------------------------------------------------------------
\begin{align}
   \left| \psi  \right\rangle & =\frac{1}{\sqrt{1+{{c}_{q=1}}^{2}+{{c}_{q=2}}^{2}}}\left( {{\left| \psi  \right\rangle }_{\mathrm{Bell}}}+{{e}^{i\pi \theta }}{{c}_{q=1}}\left| 21 \right\rangle +{{e}^{i\pi \theta }}{{c}_{q=1}}\left| 32 \right\rangle +{{e}^{i\pi \vartheta }}{{c}_{q=2}}\left| 31 \right\rangle  \right) \nonumber \\ 
 & =\frac{1}{\sqrt{1+{{c}_{q=1}}^{2}+{{c}_{q=2}}^{2}}}\left( \frac{\left| 11 \right\rangle +\left| 22 \right\rangle +\left| 33 \right\rangle }{\sqrt{3}}+{{e}^{i\pi \theta }}{{c}_{q=1}}\left| 21 \right\rangle +{{e}^{i\pi \theta }}{{c}_{q=1}}\left| 32 \right\rangle +{{e}^{i\pi \vartheta }}{{c}_{q=2}}\left| 31 \right\rangle  \right)  
\label{eq S: 3D non-maximal QFC}
\end{align}
%------------------------------------------------------------------------------------------------
Again, $c_{q=1}$ is the amplitude factor for the first upper JSI line, which comprises of the states $\left| 21 \right\rangle $ and $\left| 32 \right\rangle $ for the three-dimensional subsystem of the QFC. The second upper JSI line with $q = 2$ has the amplitude factor $c_{q=2}$ and is occupied by the state $\left| 31 \right\rangle $. The phase factors correspond to the first and the second adjacent antidiagonals are $\theta$ and $\vartheta$, respectively. For calculating the fidelity, we also included noise to the theoretical model applying the well-known formula \cite{mahmudlu2022fully_sup},\\
%------------------------------------------------------------------------------------------------
\begin{align}
\hat{\rho }=p\left| \psi  \right\rangle \left\langle  \psi  \right|+\left( 1-p \right)\frac{1}{{{d}^{2}}}\mathbb{I}
\label{eq S: noise model}
\end{align}
%------------------------------------------------------------------------------------------------
\noindent{}\textbf{Density matrix reconstruction.} We use quantum state tomography to reconstruct the density matrix from a tomographically complete set of projection measurements (quorum), described by projection wavevectors $\left| {{\psi }_{\mathrm{ }\!\!\upsilon\!\!\text{ }}} \right\rangle$ \cite{mahmudlu2022fully_sup, james2005measurement_sup}. Thereby we expect the coincidence counts to be,
%------------------------------------------------------------------------------------------------
\begin{align}
{{n}_{v}}=C\left\langle  {{\psi }_{v}} \right|\hat{\rho }\left| {{\psi }_{v}} \right\rangle 
\label{eq S: counts}
\end{align}
%------------------------------------------------------------------------------------------------
for a constant $C$ depending on the measurement time, pair generation rate, losses etc. Using our experimental data as ${{n}_{\mathrm{ }\!\!\upsilon\!\!\text{ }}}$, we reconstruct $\hat{\rho }$ using the relations \cite{ kues2017chip_sup, james2005measurement_sup}
%------------------------------------------------------------------------------------------------
\begin{align}
\hat{\rho }={{C}^{-1}}\sum\limits_{v}{{{M}_{v}}{{n}_{v}}}\\
{{M}_{v}}=\sum\limits_{x}{{{\Gamma }_{x}}}{{\left( {{B}^{-1}} \right)}_{x,v}}\\
{{B}_{x,y}}=\left\langle  {{\psi }_{x}} \right|{{\Gamma }_{y}}\left| {{\psi }_{x}} \right\rangle\\
C=\sum\limits_{k}{{{n}_{k}}\text{ for }\mathrm{tr}\left\{ {{M}_{k}} \right\}=1}
\label{eq S: QST parameters}
\end{align}
%------------------------------------------------------------------------------------------------
For $2$D tomography we applied the projections: $\left| 1 \right\rangle $, $\left| 2 \right\rangle $, $\frac{1}{\sqrt{2}}\left( \left| 1 \right\rangle +\left| 2 \right\rangle  \right)$, $\frac{1}{\sqrt{2}}\left( \left| 1 \right\rangle +i\left| 2 \right\rangle  \right)$ for the signal and the idler, respectively, resulting in $16$ projection measurements. Therefore, we set the EOM to different powers, where the amplitude of the required sideband is maximum. The modes $\left| 1 \right\rangle $ and $\left| 2 \right\rangle $ are mixed at the center between them. The photons will be collected at the central location for all projection measurements. If only one of the resonances is needed for a projection of signal or the idler, the other one is blocked by PF1. This leads to a reduction in intensity, which can be compensated by multiplying the measured coincidence counts by $2$ ($4$).

For $3$D QST, the method was extended following \cite{thew2002qudit_sup}, where the projections used are,
%------------------------------------------------------------------------------------------------
\begin{gather}
\frac{1}{\sqrt{2}}\left( \left| 1 \right\rangle +\left| 2 \right\rangle  \right)\\
\frac{1}{\sqrt{2}}\left( {{e}^{\frac{2\pi }{3}i}}\left| 1 \right\rangle +{{e}^{-\frac{2\pi }{3}i}}\left| 2 \right\rangle  \right)\\
\frac{1}{\sqrt{2}}\left( {{e}^{-\frac{2\pi }{3}i}}\left| 1 \right\rangle +{{e}^{\frac{2\pi }{3}i}}\left| 2 \right\rangle  \right)\\
\frac{1}{\sqrt{2}}\left( \left| 1 \right\rangle +\left| 3 \right\rangle  \right)\\
\frac{1}{\sqrt{2}}\left( {{e}^{\frac{2\pi }{3}i}}\left| 1 \right\rangle +{{e}^{-\frac{2\pi }{3}i}}\left| 3 \right\rangle  \right)\\
\frac{1}{\sqrt{2}}\left( {{e}^{-\frac{2\pi }{3}i}}\left| 1 \right\rangle +{{e}^{\frac{2\pi }{3}i}}\left| 3 \right\rangle  \right)\\
\frac{1}{\sqrt{2}}\left( \left| 2 \right\rangle +\left| 3 \right\rangle  \right)\\
\frac{1}{\sqrt{2}}\left( {{e}^{\frac{2\pi }{3}i}}\left| 2 \right\rangle +{{e}^{-\frac{2\pi }{3}i}}\left| 3 \right\rangle  \right)\\
\frac{1}{\sqrt{2}}\left( {{e}^{-\frac{2\pi }{3}i}}\left| 2 \right\rangle +{{e}^{\frac{2\pi }{3}i}}\left| 3 \right\rangle  \right)
\label{eq S: 3D QST parameters}
\end{gather}
%------------------------------------------------------------------------------------------------
We applied a power of $-16$ dBm ($-9$ dBm) to the EOM for the measurements while the FSRs of the QFC were \SI{50}{\giga\hertz} (\SI{100}{\giga\hertz}), at which the $0$th and the $2$nd ($4$th) band exhibited the equal intensities, which were set by comparing the count rates of these bands. By this, $\left| 1 \right\rangle $ and $\left| 3 \right\rangle $ are mixed with the frequency mode of $\left| 2 \right\rangle $, which itself remains at this position with the same intensity as the other two, respectively. Again, the surplus signal and idler modes not required at that moment were blocked by PF1, while no count correction was needed, as every projection we used consisted of exactly two modes for signal and idler, respectively.

A density matrix representing a physical quantum state has to be Hermitian and positive semi-definite. Due to experimental imperfection, this is usually not the case for the reconstructed matrix. This is treated by applying a maximal-likelihood estimation, which is a numerical optimization finding the closest physical density matrix relative to the measurement data \cite{james2005measurement_sup}.\\

\noindent{}\textbf{Fidelity.} The fidelity is a parameter that quantifies how close the measured density matrix comes to the theoretical matrix, where a unit fidelity implicates that the two states are perfectly identical. For two density matrices ${{\hat{\rho }}_{1}}$  and ${{\hat{\rho }}_{2}}$ the fidelity reads as follows
%------------------------------------------------------------------------------------------------
\begin{align}
F\left( {{{\hat{\rho }}}_{1}},{{{\hat{\rho }}}_{2}} \right)={{\left( tr\sqrt{\sqrt{{{{\hat{\rho }}}_{1}}}{{{\hat{\rho }}}_{2}}\sqrt{{{{\hat{\rho }}}_{1}}}} \right)}^{2}}
\label{eq S: Fidelity}
\end{align}
%------------------------------------------------------------------------------------------------
We calculated the fidelities from the measured quantum states with respect to the states given by Eq. (\ref{eq S: 2D non-maximal QFC}), and Eq. (\ref{eq S: 3D non-maximal QFC}) for two-dimensional and three-dimensional cases, which were mixed with noise using the formula given by Eq. (\ref{eq S: noise model}). We optimized the parameters of the amplitudes and phases of the off-diagonal JSI lines, as well as the amount of noise. The calculated fidelity $F$ is $0.975$ ($0.958$, $0.949$) for the QFC with \SI{50}{\giga\hertz} (\SI{100}{\giga\hertz}, \SI{150}{\giga\hertz}) FSR and $F = 0.697$ ($0.821$) for QFC with \SI{50}{\giga\hertz} (\SI{100}{\giga\hertz}) FSR in case of two and three dimensional tomographies, respectively.\\

\noindent{}\textbf{Entropy measures.} We used different measures to quantify the entanglement of the states reconstructed by QST. The system is maximally entangled for a value of one and there is no entanglement for a value of zero. The entanglement entropy $E\left( \psi  \right)$ is a simple measure of the entanglement of pure states $\left| \psi  \right\rangle $. It is obtained by taking the partial trace of the density matrix related to $\left| \psi  \right\rangle $ and calculating the Von Neumann entropy of the subsystem $A$ or $B$:
%------------------------------------------------------------------------------------------------
\begin{align}
E\left( {{\psi }_{\mathrm{AB}}} \right)=-tr\left( {{{\hat{\rho }}}_{\mathrm{A}}}{{\log }_{2}}{{{\hat{\rho }}}_{\mathrm{A}}} \right)=-tr\left( {{{\hat{\rho }}}_{\mathrm{B}}}{{\log }_{2}}{{{\hat{\rho }}}_{\mathrm{B}}} \right)
\label{eq S: Entanglement entropy}
\end{align}
%------------------------------------------------------------------------------------------------
For pure states, the entanglement entropy is the same for both subsystems. For mixed states, this is not generally the case and the entanglement entropy is not valid. In the experiment, however, there is always a small mixture with noise and imperfections can make pure states appear as mixed states. Nonetheless, this measure is used in the evaluation of the experiments as these effects are expected to cause only a small deviation.

For bipartite pure and mixed states, the entanglement of formation ${{E}_{\mathrm{f}}}\left( {\hat{\rho }} \right)$ can be applied, which has an explicit function of $\hat{\rho }$ for two-dimensional systems, reading 
%------------------------------------------------------------------------------------------------
\begin{align}
{{E}_{\mathrm{f}}}\left( {\hat{\rho }} \right)=h\left( \frac{1+\sqrt{1-{{C}^{2}}}}{2} \right)
\label{eq S: Entanglement formation}
\end{align}
%------------------------------------------------------------------------------------------------
with the Shannon entropy function, $h(x)=-x{{\log }_{2}}\left( x \right)-\left( 1-x \right){{\log }_{2}}\left( 1-x \right)$ and the concurrence $C$, given by $C\left( {\hat{\rho }} \right)=\text{max}\left\{ 0,{{\lambda }_{1}}-{{\lambda }_{2}}-{{\lambda }_{3}}-{{\lambda }_{4}} \right\}$ with the ${{\lambda }_{\mathrm{i}}}$ being the eigenvalues of decreasing order of the Hermitian matrix $R=\sqrt{\sqrt{{\hat{\rho }}}\tilde{\rho }\sqrt{{\hat{\rho }}}}$ with the spin-flipped density matrix $\tilde{\rho }$ . For arbitrary (i.e. mixed) high-dimensional $2$-qudit systems, there is no explicit formula of the entanglement of formation to the knowledge of the author. 
%----------------------------------Steering paper Fig. S.5.----------------------------------
\begin{figure*}[t]
    \centering
    \includegraphics[width=0.6\textwidth,trim=0mm 0mm 0mm 0mm]{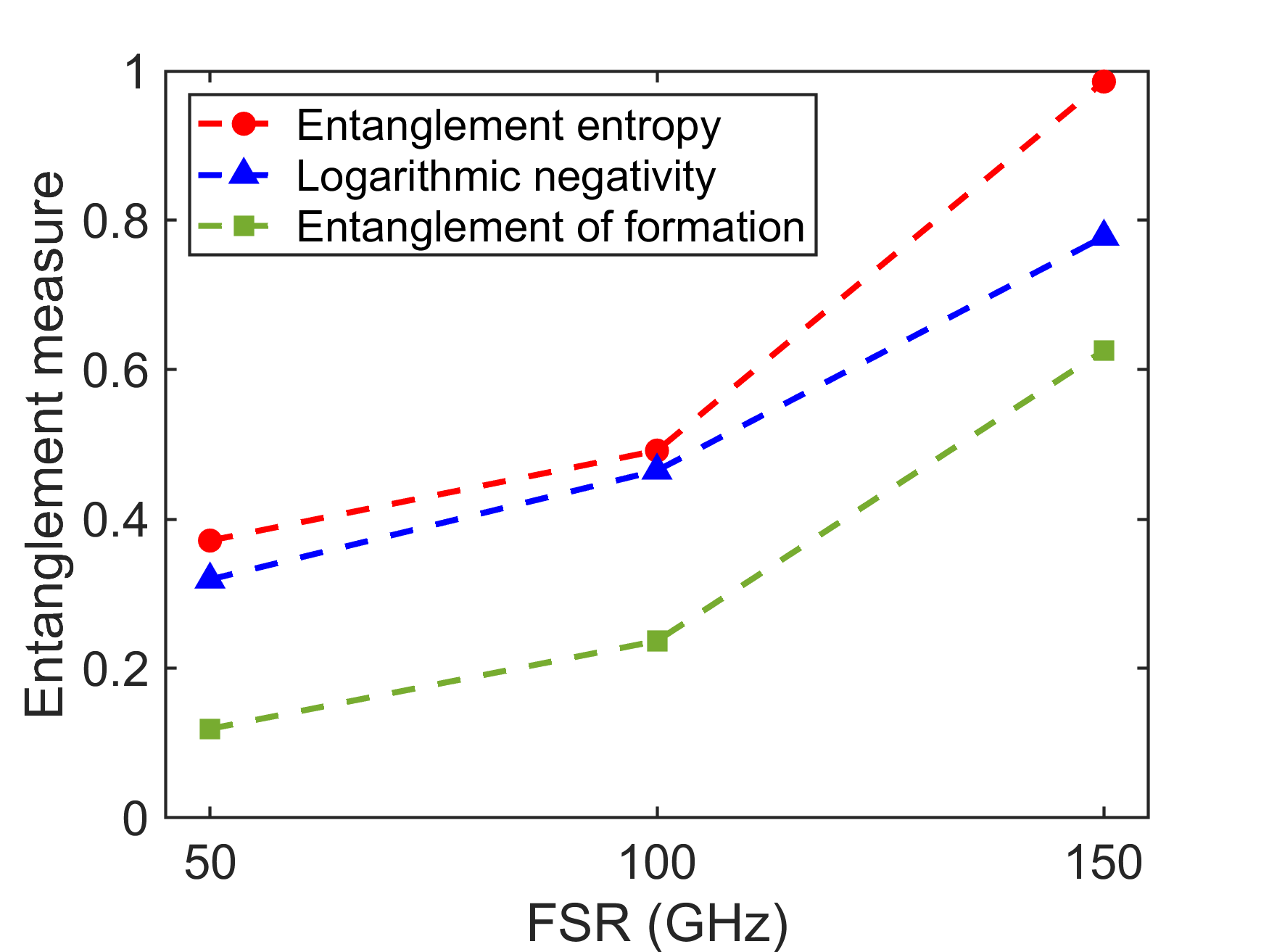}
    \caption{
    \textbf{Different parameters to measure the entanglement.} Several entanglement measures, viz. entanglement entropy, logarithmic negativity, and entanglement of formation are plotted for two-dimensional QFCs with different FSRs (\SI{50}{\giga\hertz}, \SI{100}{\giga\hertz} and \SI{150}{\giga\hertz}). The entanglement measures investigated entanglement measures were derived from the reconstructed density matrices obtained through the quantum state tomography.}
    \label{fig S: Entropy vs. FSR}
\end{figure*}
%--------------------------------Steering of QW paper Fig. S.5.----------------------------------

Another entanglement measure with an explicit formula is the logarithmic negativity, defined for two and high-dimensional pure and mixed bipartite systems as 
%------------------------------------------------------------------------------------------------
\begin{align}
{{E}_{\mathscr{{N}}}}(\hat{\rho })={{\log }_{2}}{{\left\| {{{\hat{\rho }}}^{{{T}_{\mathrm{A}}}}} \right\|}_{1}}
\label{eq S: Entanglement logarithmic negativity}
\end{align}
%-----------------------------------------------------------------------------------------------
where ${{\hat{\rho }}^{{{T}_{\mathrm{A}}}}}$ describes the partial transpose and ${{\left\| \text{ }\centerdot \text{ } \right\|}_{1}}$ denotes the trace norm.

The three presented entanglement measures are calculated for the two-dimensional density matrices for \SI{50}{\giga\hertz}, \SI{100}{\giga\hertz} and \SI{150}{\giga\hertz}, shown in Fig. S.5. All three measures show the growth of entropy with increasing FSR. The entanglement entropy reaches to almost unity for an FSR of \SI{150}{\giga\hertz}, whilst the other two measures seem to be more sensitive to noise.\\

\noindent{}\textbf{Purity.} The purity is a measure on quantum states defined as $\gamma =\text{tr}\left( {{\rho }^{2}} \right)$, which provides information on how mixed or pure a given quantum state is. If the quantum state belongs to a $d$-dimensional Hilbert space, then the purity of that state takes values within $\frac{1}{d}\le \gamma \le 1$, where $1/d$ is obtained for a completely mixed state and $\gamma =1$ describes a pure state. The purity attained from the measured density matrices correspond to the two-dimensional QFC with FSRs $\SI{150}{\giga\hertz}$, $\SI{100}{\giga\hertz}$ and $\SI{50}{\giga\hertz}$ are $85.23\%$, $85\%$, and $86\%$, respectively. Theoretically predicted Von Neumann entropies of the QFCs with FSRs $\SI{150}{\giga\hertz}$, $\SI{100}{\giga\hertz}$, and $\SI{50}{\giga\hertz}$ are $1$, $0.489$, $0.361$, respectively, which are indeed in close agreement with our experimentally obtained values. The three-dimensional two-photon states had purities of $0.435$ and $0.505$ for the FSRs of $\SI{50}{\giga\hertz}$ and $\SI{100}{\giga\hertz}$. This is significantly less than the values obtained for the two-dimensional states, nevertheless, the theoretical model of pure states, given by Eq. (\ref{eq S: 3D non-maximal QFC}) is still valid, since qutrit measurements are more prone to errors, which decreases the purity. Furthermore, note that the lower limit of the purity denoting a maximally mixed state decreases with increasing dimension.

\section{Importance of using a synchronization unit to obtain quantum walk} 
\label{Subsection: SYNC}
In this work, we also explored the importance of synchronization (SYNC) between the modulating frequency (\SI{25}{\giga\hertz}) of the EOM and the repetition rate of the fs-laser (\SI{50}{\mega\hertz}) in achieving directional quantum walk. A synchronization module (from MENLO Systems) connected to both the RF-modulator and the laser typically probes a fraction of the electrical clock signal from the RF-modulator and compares it with the laser repetition rate in real-time. A comparator and a PID controller are used to measure any discrepancy between the two and to generate a feedback signal that readjusts the laser repetition rate continuously with the RF-source by modifying the laser cavity length through a piezo-controller inside the fs-laser. For the case of bosonic and fermionic QWs initiated from MES, which were excited by a continuous-wave laser, no energy transport occurs and therefore no such synchronization is required. However, for PES, we derived that the energy transport in case of the bosonic QW can be given by Eq. (\ref{eq S: PES E-transfer}) and is proportional to $2\left| {{g}_{1}} \right|{{\Omega }_{\mathrm{FSR}}}\sin \left( {{\phi }_{\mathrm{RF}}} \right)$. Therefore, even a slight mismatch between the frequencies of the RF-modulator and the repetition rate of the fs-laser for consecutive pulse cycles resulted in an averaging of the phase fluctuations over time. This reflects in as the absence of an overall energy transport of bosonic QW. Our experiment show that a synchronization unit is indispensable to obtain directional QW.\


\begin{thebibliography}{10}%
  % <bibitems>
\expandafter\ifx\csname url\endcsname\relax
  \def\url#1{\texttt{#1}}\fi
\expandafter\ifx\csname urlprefix\endcsname\relax\def\urlprefix{URL }\fi
\providecommand{\bibinfo}[2]{#2}
\providecommand{\eprint}[2][]{\url{#2}}

%======================================================================================
% Bibitem added by Raktim for Raktim's Steering of QW paper on 19.04.2022
% Bibtex to Bibitem online converter
% https://asouqi.github.io/bibtex-converter/
%--------------------------------------------------------------------------------------
\bibitem{zhong2020quantum}Zhong, H., Wang, H., Deng, Y., Chen, M., Peng, L., Luo, Y., Qin, J., Wu, D., Ding, X., Hu, Y. \& Others Quantum computational advantage using photons. {\em Science}. \textbf{370}, 1460-1463 (2020).

\bibitem{o2009photonic}O'brien, J., Furusawa, A. \& Vučković, J. Photonic quantum technologies. {\em Nature Photonics}. \textbf{3}, 687-695 (2009).

\bibitem{erhard2020advances}Erhard, M., Krenn, M. \& Zeilinger, A. Advances in high-dimensional quantum entanglement. {\em Nature Reviews Physics}. \textbf{2}, 365-381 (2020).

\bibitem{sheridan2010security}Sheridan, L. \& Scarani, V. Security proof for quantum key distribution using qudit systems. {\em Physical Review A}. \textbf{82}, 030301 (2010).

\bibitem{cerf2002security}Cerf, N., Bourennane, M., Karlsson, A. \& Gisin, N. Security of quantum key distribution using d-level systems. {\em Physical Review Letters}. \textbf{88}, 127902 (2002).

\bibitem{kues2019quantum}Kues, M., Reimer, C., Lukens, J., Munro, W., Weiner, A., Moss, D. \& Morandotti, R. Quantum optical microcombs. {\em Nature Photonics}. \textbf{13}, 170-179 (2019).

\bibitem{kues2017chip}Kues, M., Reimer, C., Roztocki, P., Cortés, L., Sciara, S., Wetzel, B., Zhang, Y., Cino, A., Chu, S., Little, B. \& Others On-chip generation of high-dimensional entangled quantum states and their coherent control. {\em Nature}. \textbf{546}, 622-626 (2017).

\bibitem{imany201850}Imany, P., Jaramillo-Villegas, J., Odele, O., Han, K., Leaird, D., Lukens, J., Lougovski, P., Qi, M. \& Weiner, A. 50-GHz-spaced comb of high-dimensional frequency-bin entangled photons from an on-chip silicon nitride microresonator. {\em Optics Express}. \textbf{26}, 1825-1840 (2018).

\bibitem{francesconi2020engineering}Francesconi, S., Baboux, F., Raymond, A., Fabre, N., Boucher, G., Lemaitre, A., Milman, P., Amanti, M. \& Ducci, S. Engineering two-photon wavefunction and exchange statistics in a semiconductor chip. {\em Optica}. \textbf{7}, 316-322 (2020).

\bibitem{reimer2019high}Reimer, C., Sciara, S., Roztocki, P., Islam, M., Romero Cortés, L., Zhang, Y., Fischer, B., Loranger, S., Kashyap, R., Cino, A. \& Others High-dimensional one-way quantum processing implemented on d-level cluster states. {\em Nature Physics}. \textbf{15}, 148-153 (2019).

\bibitem{imany2018characterization}Imany, P., Odele, O., Jaramillo-Villegas, J., Leaird, D. \& Weiner, A. Characterization of coherent quantum frequency combs using electro-optic phase modulation. {\em Physical Review A}. \textbf{97}, 013813 (2018).

\bibitem{dakic2012quantum}Dakić, B., Lipp, Y., Ma, X., Ringbauer, M., Kropatschek, S., Barz, S., Paterek, T., Vedral, V., Zeilinger, A., Brukner, Č. \& Others Quantum discord as resource for remote state preparation. {\em Nature Physics}. \textbf{8}, 666-670 (2012)
\bibitem{agrawal2002probabilistic}Agrawal, P. \& Pati, A. Probabilistic quantum teleportation. {\em Physics Letters A}. \textbf{305}, 12-17 (2002).

\bibitem{bandyopadhyay2012optimal}Bandyopadhyay, S. \& Ghosh, A. Optimal fidelity for a quantum channel may be attained by nonmaximally entangled states. {\em Physical Review A}. \textbf{86}, 020304 (2012).

\bibitem{wang2015nonmaximally}Wang, X., Tang, S., Yuan, J. \& Kuang, L. Nonmaximally entangled states can be better for quantum correlation distribution and storage. {\em International Journal Of Theoretical Physics}. \textbf{54}, 1461-1469 (2015).

\bibitem{chen2021perfect}Chen, X., Shen, Y. \& Zhang, F. Perfect teleportation with a partially entangled quantum channel. {\em ArXiv Preprint ArXiv:2101.06693}. (2021).

\bibitem{demkowicz2006usefulness}Demkowicz-Dobrzański, R., Lewenstein, M., Sen, A., Sen, U., Bruß, D. \& Others Usefulness of classical communication for local cloning of entangled states. {\em Physical Review A}. \textbf{73}, 032313 (2006).

\bibitem{sazim2015complementarity}Sazim, S., Chakrabarty, I., Dutta, A. \& Pati, A. Complementarity of quantum correlations in cloning and deleting of quantum states. {\em Physical Review A}. \textbf{91}, 062311 (2015).

\bibitem{white1999nonmaximally}White, A., James, D., Eberhard, P. \& Kwiat, P. Nonmaximally entangled states: production, characterization, and utilization. {\em Physical Review Letters}. \textbf{83}, 3103 (1999).

\bibitem{giustina2015significant}Giustina, M., Versteegh, M., Wengerowsky, S., Handsteiner, J., Hochrainer, A., Phelan, K., Steinlechner, F., Kofler, J., Larsson, J., Abellán, C. \& Others Significant-loophole-free test of Bell’s theorem with entangled photons. {\em Physical Review Letters}. \textbf{115}, 250401 (2015).

\bibitem{zhao2016entanglement}Zhao, S., Cai, C., Liu, J., Zhou, L. \& Sheng, Y. Entanglement concentration for arbitrary four-photon cluster state assisted with single photons. {\em International Journal Of Theoretical Physics}. \textbf{55}, 1128-1144 (2016).

\bibitem{vaziri2003concentration}Vaziri, A., Pan, J., Jennewein, T., Weihs, G. \& Zeilinger, A. Concentration of higher dimensional entanglement: qutrits of photon orbital angular momentum. {\em Physical Review Letters}. \textbf{91}, 227902 (2003).

\bibitem{gomez2019experimental}G\'omez, S., Mattar, A., Machuca, I., G\'omez, E., Cavalcanti, D., Far\'ias, O., Ac\'in, A. \& Lima, G. Experimental investigation of partially entangled states for device-independent randomness generation and self-testing protocols. {\em Physical Review A}. \textbf{99}, 032108 (2019).

\bibitem{preskill2018quantum}Preskill, J. Quantum computing in the NISQ era and beyond. {\em Quantum}. \textbf{2} pp. 79 (2018).

\bibitem{wang2020qudits}Wang, Y., Hu, Z., Sanders, B. \& Kais, S. Qudits and high-dimensional quantum computing. {\em Frontiers In Physics}. pp. 479 (2020).

\bibitem{nakaji2021approximate}Nakaji, K., Uno, S., Suzuki, Y., Raymond, R., Onodera, T., Tanaka, T., Tezuka, H., Mitsuda, N. \& Yamamoto, N. Approximate amplitude encoding in shallow parameterized quantum circuits and its application to financial market indicator. {\em ArXiv Preprint ArXiv:2103.13211}. (2021).

\bibitem{mosley2008heralded}Mosley, P., Lundeen, J., Smith, B., Wasylczyk, P., U’Ren, A., Silberhorn, C. \& Walmsley, I. Heralded generation of ultrafast single photons in pure quantum states. {\em Physical Review Letters}. \textbf{100}, 133601 (2008).

\bibitem{soller2011high}Söller, C., Cohen, O., Smith, B., Walmsley, I. \& Silberhorn, C. High-performance single-photon generation with commercial-grade optical fiber. {\em Physical Review A}. \textbf{83}, 031806 (2011).

\bibitem{islam2015measuring}Islam, R., Ma, R., Preiss, P., Eric Tai, M., Lukin, A., Rispoli, M. \& Greiner, M. Measuring entanglement entropy in a quantum many-body system. {\em Nature}. \textbf{528}, 77-83 (2015).

\bibitem{harney2020entanglement}Harney, C., Pirandola, S., Ferraro, A. \& Paternostro, M. Entanglement classification via neural network quantum states. {\em New Journal Of Physics}. \textbf{22}, 045001 (2020).

\bibitem{haldar2021multi}Haldar, R., Mart\'inez, A., Mahmudlu, H., Rübeling, P., Klitis, C., Sorel, M. \& Kues, M. Multi-color Driven High-dimensional Bi-photon Quantum Frequency Combs: Versatile Joint Spectral Intensity with Tunable Entropies. {\em Frontiers In Optics}. pp. JTh5A-130 (2021).

\bibitem{haldar2021high}Haldar, R., Rübeling, P., Kashi, A. \& Kues, M. High-dimensional Bi-photon Quantum Frequency Combs with Tunable State Entropies. {\em Integrated Photonics Research, Silicon And Nanophotonics}. pp. ITh2A-5 (2021).

\bibitem{albertini2012controllability}Albertini, F. \& D’Alessandro, D. Controllability of quantum walks on graphs. {\em Mathematics Of Control, Signals, And Systems}. \textbf{24}, 321-349 (2012).

\bibitem{katayama2020floquet}Katayama, H., Hatakenaka, N. \& Fujii, T. Floquet-engineered quantum walks. {\em Scientific Reports}. \textbf{10}, 1-12 (2020).

\bibitem{novo2021floquet}Novo, L. \& Ribeiro, S. Floquet engineering of continuous-time quantum walks: Toward the simulation of complex and next-nearest-neighbor couplings. {\em Physical Review A}. \textbf{103}, 042219 (2021).

\bibitem{omar2006quantum}Omar, Y., Paunković, N., Sheridan, L. \& Bose, S. Quantum walk on a line with two entangled particles. {\em Physical Review A}. \textbf{74}, 042304 (2006).

\bibitem{haldar2021steering}Haldar, R., Kashi, A. \& Kues, M. Steering of Quantum Walks through Coherent Control of High-dimensional Bi-photon Quantum Frequency Combs. {\em The European Conference On Lasers And Electro-Optics}. pp. $1$---$1$, paper ID---cd $8.2$ (2021).

\bibitem{aharonov1993quantum}Aharonov, Y., Davidovich, L. \& Zagury, N. Quantum random walks. {\em Physical Review A}. \textbf{48}, 1687 (1993).

\bibitem{preiss2015strongly}Preiss, P., Ma, R., Tai, M., Lukin, A., Rispoli, M., Zupancic, P., Lahini, Y., Islam, R. \& Greiner, M. Strongly correlated quantum walks in optical lattices. {\em Science}. \textbf{347}, 1229-1233 (2015).

\bibitem{schreiber2010photons}Schreiber, A., Cassemiro, K., Potoček, V., Gábris, A., Mosley, P., Andersson, E., Jex, I. \& Silberhorn, C. Photons walking the line: a quantum walk with adjustable coin operations. {\em Physical Review Letters}. \textbf{104}, 050502 (2010).

\bibitem{qiang2016efficient}Qiang, X., Loke, T., Montanaro, A., Aungskunsiri, K., Zhou, X., O’Brien, J., Wang, J. \& Matthews, J. Efficient quantum walk on a quantum processor. {\em Nature Communications}. \textbf{7}, 1-6 (2016).

\bibitem{childs2009universal}Childs, A. Universal computation by quantum walk. {\em Physical Review Letters}. \textbf{102}, 180501 (2009).

\bibitem{lovett2010universal}Lovett, N., Cooper, S., Everitt, M., Trevers, M. \& Kendon, V. Universal quantum computation using the discrete-time quantum walk. {\em Physical Review A}. \textbf{81}, 042330 (2010).

\bibitem{zhang2016creating}Zhang, W., Goyal, S., Gao, F., Sanders, B. \& Simon, C. Creating cat states in one-dimensional quantum walks using delocalized initial states. {\em New Journal Of Physics}. \textbf{18}, 093025 (2016).

\bibitem{mohseni2008environment}Mohseni, M., Rebentrost, P., Lloyd, S. \& Aspuru-Guzik, A. Environment-assisted quantum walks in photosynthetic energy transfer. {\em The Journal Of Chemical Physics}. \textbf{129}, 11B603 (2008).

\bibitem{schuld2014quantum}Schuld, M., Sinayskiy, I. \& Petruccione, F. Quantum walks on graphs representing the firing patterns of a quantum neural network. {\em Physical Review A}. \textbf{89}, 032333 (2014).

\bibitem{ming2019quantum}Ming, Y., Lin, C., Bartlett, S. \& Zhang, W. Quantum topology identification with deep neural networks and quantum walks. {\em Npj Computational Materials}. \textbf{5}, 1-7 (2019).

\bibitem{hameroff2014quantum}Hameroff, S. Quantum walks in brain microtubules—A biomolecular basis for quantum cognition?. {\em Topics In Cognitive Science}. \textbf{6}, 91-97 (2014).

\bibitem{peruzzo2010quantum}Peruzzo, A., Lobino, M., Matthews, J., Matsuda, N., Politi, A., Poulios, K., Zhou, X., Lahini, Y., Ismail, N., Wörhoff, K. \& Others Quantum walks of correlated photons. {\em Science}. \textbf{329}, 1500-1503 (2010).

\bibitem{cardano2017detection}Cardano, F., D’Errico, A., Dauphin, A., Maffei, M., Piccirillo, B., Lisio, C., De Filippis, G., Cataudella, V., Santamato, E., Marrucci, L. \& Others Detection of Zak phases and topological invariants in a chiral quantum walk of twisted photons. {\em Nature Communications}. \textbf{8}, 1-7 (2017).

\bibitem{sansoni2012two}Sansoni, L., Sciarrino, F., Vallone, G., Mataloni, P., Crespi, A., Ramponi, R. \& Osellame, R. Two-particle bosonic-fermionic quantum walk via integrated photonics. {\em Physical Review Letters}. \textbf{108}, 010502 (2012).

\bibitem{imany2020probing}Imany, P., Lingaraju, N., Alshaykh, M., Leaird, D. \& Weiner, A. Probing quantum walks through coherent control of high-dimensionally entangled photons. {\em Science Advances}. \textbf{6}, eaba8066 (2020).

\bibitem{panahiyan2018controlling}Panahiyan, S. \& Fritzsche, S. Controlling quantum random walk with a step-dependent coin. {\em New Journal Of Physics}. \textbf{20}, 083028 (2018).

\bibitem{de2010tailoring}Valcárcel, G., Roldán, E. \& Romanelli, A. Tailoring discrete quantum walk dynamics via extended initial conditions. {\em New Journal Of Physics}. \textbf{12}, 123022 (2010).

\bibitem{dadras2018quantum}Dadras, S., Gresch, A., Groiseau, C., Wimberger, S. \& Summy, G. Quantum walk in momentum space with a Bose-Einstein condensate. {\em Physical Review Letters}. \textbf{121}, 070402 (2018).

\bibitem{weiss2015steering}Weiß, M., Groiseau, C., Lam, W., Burioni, R., Vezzani, A., Summy, G. \& Wimberger, S. Steering random walks with kicked ultracold atoms. {\em Physical Review A}. \textbf{92}, 033606 (2015).

\bibitem{kashi2022frequency}Kashi, A., Sader, L., Haldar, R., Wetzel, B. \& Kues, M. Frequency-to-time mapping technique for direct spectral characterization of biphoton states from pulsed spontaneous parametric processes. {\em Frontiers In Photonics}. \textbf{3} pp. 834065 (2022).

\bibitem{khodadad2021spectral}Khodadad Kashi, A. \& Kues, M. Spectral Hong–Ou–Mandel Interference between Independently Generated Single Photons for Scalable Frequency-Domain Quantum Processing. {\em Laser \& Photonics Reviews}. \textbf{$15$}, $2000464$ ($2021$).

\bibitem{u2006generation}U’Ren, A., Erdmann, R., La Cruz-Gutierrez, M. \& Walmsley, I. Generation of two-photon states with an arbitrary degree of entanglement via nonlinear crystal superlattices. {\em Physical Review Letters}. \textbf{97}, 223602 (2006).

\bibitem{kumar2014controlling}Kumar, R., Ong, J., Savanier, M. \& Mookherjea, S. Controlling the spectrum of photons generated on a silicon nanophotonic chip. {\em Nature Communications}. \textbf{5}, 1-7 (2014).

\bibitem{leefmans2022topological}Leefmans, C., Dutt, A., Williams, J., Yuan, L., Parto, M., Nori, F., Fan, S. \& Marandi, A. Topological dissipation in a time-multiplexed photonic resonator network. {\em Nature Physics}. pp. 1-8 (2022).

\bibitem{wang2021generating}Wang, K., Dutt, A., Yang, K., Wojcik, C., Vučković, J. \& Fan, S. Generating arbitrary topological windings of a non-Hermitian band. {\em Science}. \textbf{371}, 1240-1245 (2021).

\bibitem{wergen2011record}Wergen, G., Bogner, M. \& Krug, J. Record statistics for biased random walks, with an application to financial data. {\em Physical Review E}. \textbf{83}, 051109 (2011).

\bibitem{codling2008random}Codling, E., Plank, M. \& Benhamou, S. Random walk models in biology. {\em Journal Of The Royal Society Interface}. \textbf{5}, 813-834 (2008).\\

%====================================Steering of QW PPLN References Main Manuscript===================================================


\end{thebibliography}

\begin{thebibliography}{10}
\expandafter\ifx\csname url\endcsname\relax
  \def\url#1{\texttt{#1}}\fi
\expandafter\ifx\csname urlprefix\endcsname\relax\def\urlprefix{URL }\fi
\providecommand{\bibinfo}[2]{#2}
\providecommand{\eprint}[2][]{\url{#2}}

% Previous references used in the main manuscript -----------------------------------------
\bibitem{haldar2021high_sup}Haldar, R., Rübeling, P., Kashi, A. \& Kues, M. High-dimensional Bi-photon Quantum Frequency Combs with Tunable State Entropies. {\em Integrated Photonics Research, Silicon And Nanophotonics}. pp. ITh2A-5 (2021).

\bibitem{haldar2021multi_sup}Haldar, R., Mart\'inez, A., Mahmudlu, H., R\"ubeling, P., Klitis, C., Sorel, M. \& Kues, M. Multi-color Driven High-dimensional Bi-photon Quantum Frequency Combs: Versatile Joint Spectral Intensity with Tunable Entropies. {\em Frontiers In Optics}. pp. JTh5A-130 (2021).

\bibitem{kumar2014controlling_sup}Kumar, R., Ong, J., Savanier, M. \& Mookherjea, S. Controlling the spectrum of photons generated on a silicon nanophotonic chip. {\em Nature Communications}. \textbf{5}, 1-7 (2014).

\bibitem{kues2017chip_sup}Kues, M., Reimer, C., Roztocki, P., Cortés, L., Sciara, S., Wetzel, B., Zhang, Y., Cino, A., Chu, S., Little, B. \& Others On-chip generation of high-dimensional entangled quantum states and their coherent control. {\em Nature}. \textbf{546}, 622-626 (2017).


\bibitem{imany2020probing_sup}Imany, P., Lingaraju, N., Alshaykh, M., Leaird, D. \& Weiner, A. Probing quantum walks through coherent control of high-dimensionally entangled photons. {\em Science Advances}. \textbf{6}, eaba8066 (2020).

% New references added in the Supplementary Information -----------------------------------
\bibitem{hu2020realization_sup}Hu, Y., Reimer, C., Shams-Ansari, A., Zhang, M. \& Loncar, M. Realization of high-dimensional frequency crystals in electro-optic microcombs. {\em Optica}. \textbf{7}, 1189-1194 (2020).

\bibitem{capmany2010quantum_sup}Capmany, J. \& Fernández-Pousa, C. Quantum model for electro-optical phase modulation. {\em JOSA B}. \textbf{27}, A119-A129 (2010).

\bibitem{rueda2019resonant_sup}Rueda, A., Sedlmeir, F., Kumari, M., Leuchs, G. \& Schwefel, H. Resonant electro-optic frequency comb. {\em Nature}. \textbf{568}, 378-381 (2019).

\bibitem{mahmudlu2022fully_sup}Mahmudlu, H., Johanning, R., Kashi, A., Rees, A., Epping, J., Haldar, R., Boller, K. \& Kues, M. Fully on-chip photonic turnkey quantum source for entangled qubit/qudit state generation. {\em ArXiv Preprint ArXiv:2206.08715}. (2022).

\bibitem{james2005measurement_sup}James, D., Kwiat, P., Munro, W. \& White, A. On the measurement of qubits. {\em Asymptotic Theory Of Quantum Statistical Inference: Selected Papers}. pp. 509-538 (2005).

\bibitem{thew2002qudit_sup}Thew, R., Nemoto, K., White, A. \& Munro, W. Qudit quantum-state tomography. {\em Physical Review A}. \textbf{66}, 012303 (2002).


\end{thebibliography}
\end{document}